\documentclass[superscriptaddress,prb,twocolumn,footinbib]{revtex4-2}
\usepackage[bookmarks=false,colorlinks=true,urlcolor=blue,citecolor=blue,linkcolor=blue]{hyperref}

\usepackage{xspace}
\usepackage{tikz}
\usetikzlibrary{calc,arrows}

\usepackage{amsmath,amssymb}
\usepackage{gensymb}
\usepackage[normalem]{ulem}
\usepackage{booktabs}

\newcommand{\mycomment}[1]{}

\newcommand{\markup}[1]{#1}

\newcommand{\delete}[1]{}

\newcommand{\add}[1]{#1}

\newcommand{\para}{{\mkern3mu\vphantom{\perp}\vrule depth 0pt\mkern2mu\vrule depth 0pt\mkern3mu}}

\newcommand{\be}{\begin{equation}}
\newcommand{\ee}{\end{equation}}
\newcommand{\bea}{\begin{eqnarray}}
\newcommand{\eea}{\end{eqnarray}}
\newcommand{\beal}{\begin{align}}
\newcommand{\eeal}{\end{align}}
\newcommand{\bes}{\begin{equation} \begin{split}}
\newcommand{\ees}{\end{split} \end{equation}}

\newcommand{\f}{\frac}

\newcommand{\qv}{\vec{q}}

\newcommand{\Qv}{\vec{Q}}

\newcommand{\rv}{\vec{r}}
\newcommand{\rvp}{\vec{r}^{\,\prime}}
\newcommand{\Rv}{\vec{R}}

\newcommand{\Gv}{\vec{G}}

\newcommand{\uv}{\vec{u}}
\newcommand{\vv}{\vec{v}}

\newcommand{\wv}{\vec{w}}

\newcommand{\Sv}{\vec{S}}

\newcommand{\av}{\vec{a}}
\newcommand{\bv}{\vec{b}}
\newcommand{\mv}{\vec{m}}
\newcommand{\hv}{\vec{h}}

\newcommand{\SLPONE}{SLP$_{1}$\xspace}
\newcommand{\SLPTWO}{SLP$_{2}$\xspace}
\newcommand{\SLPTHREE}{SLP$_{3}$\xspace}
\newcommand{\SLPTHREEM}{SLP$_{3,M}$\xspace}
\newcommand{\SLPTHREEALPHA}{SLP$_{3,\alpha}$\xspace}
\newcommand{\SLPTHREECHIZERO}{SLP$_{3,\chi=0}$\xspace}

\newcommand{\RN}[1]{%
\textup{\uppercase\expandafter{\romannumeral#1}}%
}

\usepackage{contour}

\definecolor{taylorswift}{rgb}{0.0862745098,0.4666666667,0.3411764706}
\definecolor{fearless}{rgb}{0.8862745098,0.6117647059,0.2823529412}
\definecolor{speaknow}{rgb}{0.4588235294,0.2274509804,0.4980392157}
\definecolor{red}{rgb}{0.6509803922,0.1254901961,0.2705882353}
\definecolor{TS1989}{rgb}{0.1803921569,0.6,0.9764705882}
\definecolor{reputation}{rgb}{0.1450980392,0.1490196078,0.1529411765}
\definecolor{lover}{rgb}{0.8392156863,0.2117647059,0.5529411765}

\tikzset{>=latex}

\begin{document}

\title{Field-induced Multi-$\boldsymbol{\vec{Q}}$ States in a Pyrochlore Heisenberg Magnet}

\author{Cecilie Glittum}
\affiliation{T.C.M. Group, Cavendish Laboratory, JJ Thomson Avenue, Cambridge CB3 0HE, United Kingdom}
\affiliation{Helmholtz-Zentrum Berlin für Materialien und Energie GmbH, Hahn-Meitner-Platz 1 14109 Berlin, Germany}
\affiliation{Dahlem Center for Complex Quantum Systems and Fachbereich Physik, Freie Universität Berlin, 14195 Berlin, Germany}
\author{Olav F. Sylju{\aa}sen}
\affiliation{Department of Physics, University of Oslo, P.~O.~Box 1048 Blindern, N-0316 Oslo, Norway}

\begin{abstract}
We construct exact ground states of the $J_1$-$J_{3b}$ classical Heisenberg model on the pyrochlore lattice in the presence of a magnetic field. They are noncoplanar multi-$\Qv$ spin configurations with a large magnetic unit cell that generalize the previously found coplanar sublattice pairing states.
Using linear spin-wave theory, we show that entropy favors these multi-$\Qv$ states at low temperatures in high magnetic fields. This is confirmed by Monte Carlo simulations, and a phase diagram is constructed. We also calculate the zero-temperature dynamical structure factor. Besides the usual Goldstone modes associated with the ordering $\Qv$s, we find high intensity gapless modes at momenta where there are no Bragg peaks.
\end{abstract}

\date{\today}

\maketitle

\section{Introduction \label{sec:introduction}}
\noindent

Frustrated magnets, and particularly the highly frustrated pyrochlores, have garnered significant attention due to their predicted ability to host exotic states of matter, such as classical spin liquids. The antiferromagnetic (AF) nearest-neighbor Heisenberg model on the pyrochlore lattice is such a spin liquid \delete{, even down to zero temperature,} due to its huge ground state degeneracy~\cite{Villain1979,Reimers1991,Reimers1992,Moessner1998}. This degeneracy is however lifted by further-neighbor couplings, and inclusion of second- and third-nearest neighbor couplings destabilize the spin liquid, inducing different kinds of order~\cite{Chern2008, Okubo2011, Zhitomirsky2022, Glittum2023}.

While the primary interest in frustrated magnetism often lies in the lack of magnetic order in presence of strong correlations, i.e. spin liquidity, noncoplanar ordered states with several ordering wave vectors have also gained much attention, particularly due to their applications in spintronics~\cite{Nagaosa2013, Koshibae2015, Fujishiro2018}.
Such multi-$\Qv$ states are typically known to be realized in the presence of the Dzyaloshinskii–Moriya interaction~\cite{Muhlbauer2009,Yi2009, Buhrandt2013}. However, they can also be stabilized by frustration~\cite{Okubo2012},  particularly with additional anisotropic exchange interactions~\cite{Leonov2015, Hayami2016_1}, impurities~\cite{Hayami2016_2} or coupling to itinerant electrons~\cite{Martin2008,Ozawa2016, Hayami2017}. These mechanisms are also known to induce multi-$\Qv$ states on pyrochlore lattices~\cite{Chern2008,Okubo2011,Chern2010,Zhitomirsky2022, Aoyama2021,Aoyama2022}.

\begin{figure}[h!]
\begin{center}
\includegraphics[width=\columnwidth]{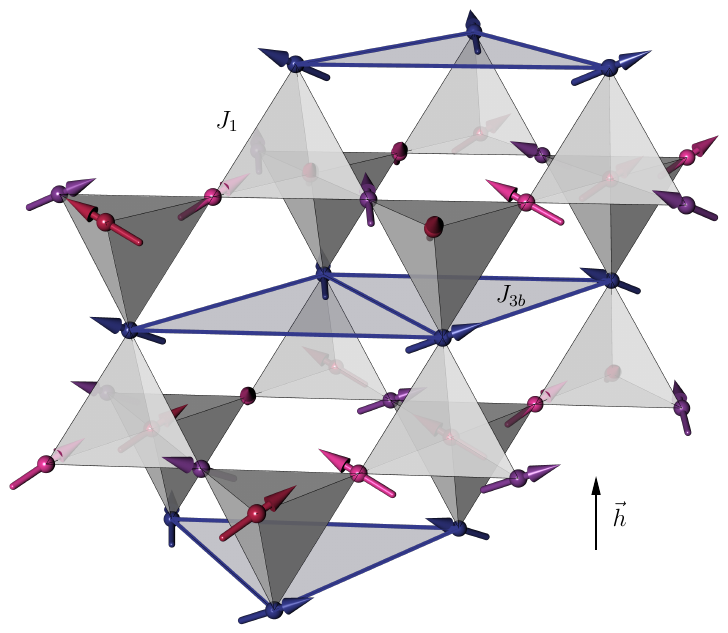}
\caption{Pyrochlore lattice with an example of the canted \SLPONE state in a magnetic field $h=4$.
On sublattice 1 (blue), we illustrate how $J_{3b}$ effectively couples spins in triangular planes.
In the canted \SLPONE state, spins on pairs of sublattices form antiparallel spirals in the $\perp$-components (relative to the magnetic field direction), while all $\para$-components are equal. Each triangular plane on a given sublattice has the same three spins.
 The \SLPONE state can also be viewed as chains of spins with staggered $\perp$-components -- here for chains of pink and blue spins and chains of red and purple spins.
 }
\label{fig:SLP1}
\end{center}
\end{figure}

The third nearest-neighbor couplings are predicted to be the most important further-neighbor couplings in several pyrochlore materials~\cite{Wills2006, Yaresko2008,Cheng2008}. For the pure AF $J_1$-$J_{3b}$ Heisenberg model (see Fig.~\ref{fig:SLP1}), coplanar sublattice pairing (SLP) states are known to minimize the $J_1$ and $J_{3b}$ energy terms simultaneously~\cite{Glittum2023}. In these states, sublattices pair up antiferromagnetically, forming coplanar single-$\Qv$ spirals corresponding to $120\degree$ order in the triangular planes formed by the $J_{3b}$ coupling.

\markup{Here, we study the impact of an applied magnetic field on the pyrochlore AF $J_1$-$J_{3b}$ Heisenberg model. We start in Sec.~\ref{sec:Hamiltonian} by constructing an ansatz for the system's ground state as a generalized magnetic-field analog of the SLP ansatz of Ref.~\onlinecite{Glittum2023}. This ansatz allows for a simultaneous SLP between all pairs of sublattices, and can describe more general states than the simple single-$\Qv$ spiral on each sublattice. Instead, each sublattice has up to three ordering wave vectors, one for each sublattice pair. The system thus has up to six ordering wave vectors in total.
In Sec.~\ref{sec:solutions}, we present normalized multi-$\Qv$ states satisfying the ansatz, and show further in Sec.~\ref{sec:spinwaveentropy} that $6\Qv$-states are favored entropically at sufficiently high magnetic fields.
We define order parameters measuring various aspects of SLP order in Sec.~\ref{sec:orderparameters}, and confirm our findings numerically using Monte Carlo (MC) simulations in Sec.~\ref{sec:results}, where we also map out the phase diagram. We compute zero-temperature static structure factors in Sec.~\ref{sec:staticsf} and dynamical structure factors, i.e., spin waves, in Sec.~\ref{sec:dynamicalsf} for selected values of the magnetic field, and end with a discussion of our findings in Sec.~\ref{sec:discussion}.}

\section{Hamiltonian \label{sec:Hamiltonian}}

We consider the following classical Heisenberg Hamiltonian in a magnetic field $\hv$ on the pyrochlore lattice
\begin{align}
  H &= J_1 \! \! \sum_{\langle \rv,\rvp \rangle}  \! \! \Sv_{\rv} \cdot \Sv_{\rvp}  + J_{3b} \! \! \! \! \! \! \sum_{\langle \langle \langle \rv,\rvp \rangle \rangle \rangle_b} \! \! \! \! \! \! \Sv_{\rv} \cdot \Sv_{\rvp}  - \sum_{\rv} \hv\cdot \Sv_{\rv},
\end{align}
where the AF exchange couplings $J_1=1$ and $J_{3b} > 0$ are illustrated in Fig.~\ref{fig:SLP1}.

We begin by rewriting the Hamiltonian as
\begin{equation}
\begin{split}
 H &= \f{J_1}{2} \sum_{t}  \left( \Sv_{t,0} + \Sv_{t,1} + \Sv_{t,2} + \Sv_{t,3} - 4\mv \right)^2 \label{eq:Hamiltonian2}\\
 & + \f{J_{3b}}{4} \sum_{s=0}^{3} \sum_{\triangle} \left( \Sv_{i_0,s}+\Sv_{i_1,s} + \Sv_{i_2,s} - 3\mv \right)^2,
\end{split}
\end{equation}
where the first sum goes over all up and down tetrahedra $t$ of the lattice, of which there are $N/2$ ($N$ is the number of spins), and $\Sv_{t,s}$ denotes the spin on tetrahedron $t$ and sublattice $s$.  The second sum goes over all elementary triangles on the triangular lattices formed by the $J_{3b}$ bonds, which couple spins on the same sublattice $s$ (see Fig.~\ref{fig:SLP1}). The number of such triangles on each sublattice is also $N/2$. The sites on each triangle are denoted $i_0,i_1$ and $i_2$.
%The factors $1/2$ in front of each term is to compensate for the factors of 2 in multiplying out the parentheses.
%The extra factor of $1/2$ in the last term is to account for the fact that a given $J_{3b}$ interaction between two spins occurs in two triangles.
We have subtracted a constant term coming from the spins and the average magnetization squared ($\mv$ is the average magnetization per spin).
Multiplying out the parentheses, the resulting Hamiltonian is that of the $J_1$-$J_{3b}$ model in a magnetic field, where the magnetic field is along the axis of the magnetization and has magnitude
$h    =  \left( 8J_1 + 9J_{3b} \right) m$.
This gives a saturation field $h_{\rm sat} = 8J_1 + 9J_{3b}$, and an average magnetization per spin
\be
m = \f{h}{8 J_1 + 9 J_{3b}}. \label{eq:magnetization}
\ee
The ground state energy (the constant we subtracted earlier) per spin is
\be
   \f{E_{\rm GS}}{N} = - \left( J_1 + \f{3}{2} J_{3b} + \f{h^2}{2(8J_1 + 9 J_{3b})} \right). \label{eq:groundstateenergy}
\ee

In Eq.~\eqref{eq:Hamiltonian2}, we have chosen to write the Hamiltonian so that the lowest energy is achieved when the spins satisfy
 \be
\Sv_{t,0} + \Sv_{t,1} + \Sv_{t,2} + \Sv_{t,3} = 4\mv \label{tetrahedroncondition}
 \ee
for each tetrahedron and
\be
       \Sv_{i_0,s}+\Sv_{i_1,s} + \Sv_{i_2,s} =  3\mv \label{trianglecondition}
 \ee
 for each triangle. We will refer to Eq.~\eqref{tetrahedroncondition} (Eq.~\eqref{trianglecondition}) as the tetrahedron (triangle) condition. The rationale behind this choice is to match it with known results in the cases $J_1>0, J_{3b}= 0$~\cite{Penc2004} and $J_1=0$,$J_{3b}>0$~\cite{Kawamura1985}.

\subsection{Ansatz}
In the following, we label the spins by either their face-centered cubic (fcc) lattice point $\Rv$ and sublattice index $i=\{0,1,2,3\}$, or their position \markup{$\rv = \Rv + \vec{\delta}_i$.} $\Rv$ describes the position of an up tetrahedron and is constructed from the fcc primitive lattice vectors $\vec{a}_1 = (0,1/2,1/2), \vec{a}_2 = (1/2,0,1/2)$,  and $\vec{a}_3 = (1/2,1/2,0)$, where we have set the cubic lattice constant to unity. The corresponding reciprocal lattice vectors are $\bv_1= 2\pi(-1,1,1)$, $\bv_2= 2\pi(1,-1,1)$, and $\bv_3= 2\pi(1,1,-1)$. \markup{The sublattice vectors $\vec{\delta}_i$, which describe the spin positions on a single up tetrahedron, are $\vec{\delta}_{i} = \vec{a}_i/2$ for $i=\{0,1,2,3 \}$, where we have defined $\vec{a}_0=(0,0,0)$.}

A general ansatz satisfying the tetrahedron and triangle conditions is
\begin{equation}\label{eq:general ansatz}
\begin{split}
  \Sv_{\Rv, 0} = &+ \uv_{01} \cos{(\Qv_{01} \cdot \Rv)}+ \vv_{01} \sin{(\Qv_{01} \cdot \Rv)}\\
  & +\uv_{02} \cos{(\Qv_{02} \cdot \Rv)}+ \vv_{02} \sin{(\Qv_{02} \cdot \Rv)}\\
  & +\uv_{03} \cos{(\Qv_{03} \cdot \Rv)}+ \vv_{03} \sin{(\Qv_{03} \cdot \Rv)} + \mv, \\
  \Sv_{\Rv, 1} = &-\uv_{01} \cos{(\Qv_{01} \cdot \Rv)}- \vv_{01} \sin{(\Qv_{01} \cdot \Rv)}\\
  & +\uv_{12} \cos{(\Qv_{12} \cdot \Rv)}+ \vv_{12} \sin{(\Qv_{12} \cdot \Rv)}\\
  & +\uv_{13} \cos{(\Qv_{13} \cdot \Rv)}+ \vv_{13} \sin{(\Qv_{13} \cdot \Rv)} + \mv,\\
  \Sv_{\Rv, 2} = &-\uv_{02} \cos{(\Qv_{02} \cdot \Rv)}- \vv_{02} \sin{(\Qv_{02} \cdot \Rv)}\\
  & -\uv_{12} \cos{(\Qv_{12} \cdot \Rv)}- \vv_{12} \sin{(\Qv_{12} \cdot \Rv)}\\
  & +\uv_{23} \cos{(\Qv_{23} \cdot \Rv)}+ \vv_{23} \sin{(\Qv_{23} \cdot \Rv)} + \mv,\\
  \Sv_{\Rv, 3} = &-\uv_{03} \cos{(\Qv_{03} \cdot \Rv)}- \vv_{03} \sin{(\Qv_{03} \cdot \Rv)}\\
  & -\uv_{13} \cos{(\Qv_{13} \cdot \Rv)}- \vv_{13} \sin{(\Qv_{13} \cdot \Rv)}\\
  & -\uv_{23} \cos{(\Qv_{23} \cdot \Rv)}- \vv_{23} \sin{(\Qv_{23} \cdot \Rv)} + \mv,
\end{split}
\end{equation}
where the wave vectors $\Qv_{ij}$ are listed in Table~\ref{Qijs}. Note that the minus signs in the ansatz cause an antiparallel tendency of spins on pairs of sublattices, hence we term this ansatz the SLP ansatz.

\begin{table}
  \caption{The $\Qv_{ij}$s in terms of reciprocal lattice vectors. $\Rv=n_1 \av_{1} + n_2 \av_{2} + n_3 \av_{3}$.\\
    \label{Qijs}}
  \begin{tabular}{ccc}
\toprule
$\Qv_{ij}$ & momentum & $\Qv \cdot \Rv$\\
\hline
$\Qv_{01}$ &  $(\bv_3-\bv_2)/3$              & $(2\pi/3) \left( n_3-n_2 \right)$\\
$\Qv_{02}$ & $(\bv_1-\bv_3)/3$               & $(2\pi/3) \left( n_1-n_3 \right)$\\
$\Qv_{03}$ & $(\bv_2-\bv_1)/3$               & $(2\pi/3) \left( n_2-n_1 \right)$\\
$\Qv_{23}$ & $(2\bv_1+\bv_2+\bv_3)/3$ & $(2\pi/3) \left( 2n_1+n_2+n_3 \right)$\\
$\Qv_{13}$ & $(\bv_1+2\bv_2+\bv_3)/3$ & $(2\pi/3) \left( n_1 +2n_2+n_3 \right)$\\
$\Qv_{12}$ & $(\bv_1+\bv_2+2\bv_3)/3$ & $(2\pi/3) \left( n_1 + n_2 + 2n_3 \right)$\\
\bottomrule
\end{tabular}
\end{table}
The choice of minus signs gives  $\Sv_{\Rv, 0} +\Sv_{\Rv, 1} +\Sv_{\Rv, 2} + \Sv_{\Rv, 3}=4\mv$, meaning that the tetrahedron condition is satisfied for all up tetrahedra.
The tetrahedron condition is also satisfied for the down tetrahedra, $\sum_{i} \Sv_{\Rv-\av_i,i}= 4 \mv$.
This can be seen by letting $\Rv \to \Rv - \av_i$ on sublattice $i$. It gives the same cancellations as for the up tetrahedra, which follows from $\Qv_{ij} \cdot \av_i = \Qv_{ij} \cdot \av_j$, as one can easily check by inspecting Table~\ref{Qijs}.

To see that also the triangle condition is satisfied, it is necessary to divide each sublattice into parallel triangular $J_{3b}$ planes (as shown for one of the sublattices in Fig.~\ref{fig:SLP1}), and furthermore, divide each triangular plane into trisublattices, such that an elementary triangle has one site from each trisublattice. A possible way of doing this is to introduce integer-valued plane and trisublattice indices, $P$ and $T$, respectively, as defined in Table~\ref{LTs}. When rewriting the SLP ansatz in Eq.~\eqref{eq:general ansatz} with this labeling, we find that the particular choices of $\Qv_{ij}$s in Table~\ref{Qijs} cause all trigonometric functions to have arguments which depend on the trisublattice index as $2\pi T/3$ and on the plane index as $2\pi P/3$. Therefore, for each elementary triangle in a given plane,
%where $T$ takes on the three different values $\{0,1,2\}$,
the trigonometric functions from the different trisublattices cancel each other, and the triangle condition is satisfied. We also note that there are at most three different planes on a given sublattice, which follows from the $2\pi P/3$ dependence, making the spin configuration periodic when $P$ changes by 3.
\begin{table}
  \caption{
The triangular plane basis vectors $\vec{t}_1,\vec{t}_2$ and their ``normal'' vector $\vec{n}$ on the different sublattices, as well as their plane index $P$ and trisublattice index $T$.
%The triangular basis vectors are chosen such that they differ by 120\degree and form a positive chirality quantity with the ``normal'' vector $\vec{n} \cdot (\vec{t}_1 \times \vec{t}_2) >0$, and the ``normal'' vector defines the vector between the same trisublattice sites on neighboring parallel planes.
A general point $\Rv = n_1 \av_1 + n_2 \av_2 + n_3 \av_3 = P \vec{n} + m_1 \vec{t}_1 + m_2 \vec{t}_2 $, and $T \equiv (m_1+m_2)\mod 3$. We have defined $n_0 = -n_1-n_2-n_3$. $T$ is understood to be mod 3.
\label{LTs}}
%\begin{tabular*}{\textwidth}{cccccc}
\begin{tabular}{cccccc}
\toprule
sublattice &$\vec{t}_1$    & $\vec{t}_2$     & $\vec{n}$ & $T$ & $P$ \\
\midrule
0 &  $-\av_3+\av_1$          & $-\av_1+\av_2$      & $\av_1-\av_0$  & $n_2-n_3$ & $-n_0$\\
1 &  $+\av_2-\av_0$          &  $+\av_0-\av_3$     & $\av_0-\av_1$  & $n_2-n_3$ & $-n_1$\\
2 &  $-\av_1+\av_3$          &  $-\av_3+\av_0$     & $\av_3-\av_2$  & $n_0-n_1$ & $-n_2$\\
3 &  $+\av_0-\av_2$          & $+\av_2-\av_1$      & $\av_2-\av_3$  & $n_0-n_1$ & $-n_3$\\
\bottomrule
\end{tabular}
\end{table}

%For ordinary SLP, where only $\uv_{01},\vv_{01},\uv_{23}$ and $\vv_{23}$ are nonzero, we see that this ansatz requires the nonzero $\uv$'s and $\vv$'s to be perpendicular to $\mv$ as otherwise the length of spins cannot be unity. This implies that all spin-z components are the same, and we have the canted SLP state. Note that the length of the nonzero $\uv$'s and $\vv$'s will depend on $m$. In particularly they go to 0 as $m \to 1$.
%If the $\uv$ and $\vv$ are allowed to have nonzero z-components, the spins will also have different z-components as well. Whether or not such a state is possible depends on the existence of a solution of the 36 length constraint equations.

\section{Ground states \label{sec:solutions}}
Since the SLP ansatz in Eq.~\eqref{eq:general ansatz} fulfills both the tetrahedron and triangle conditions, any proper state satisfying the ansatz is a ground state of the Hamiltonian. To be a proper classical spin state, each spin must have unit length.  It is in general difficult to achieve this for multi-$\Qv$ states, as they often have a large number of different spins in the unit cell and a corresponding large number of normalization constraints. Here, the $\Qv_{ij}$s make any spin configuration that obeys Eq.~\eqref{eq:general ansatz} periodic with unit cell $(3\av_1,3\av_2,3\av_3)$, with a total of 108 spins. However, only 36 of these spins can be different, nine on each sublattice (at most three different planes with three spins each on each sublattice),  giving rise to 36 normalization constraints. These are listed in Appendix~\ref{app:lengthconstraints}.

The general SLP ansatz, Eq.~\eqref{eq:general ansatz}, allows for several solutions that fulfill the length constraints. It is not the purpose here to find all solutions, but rather to describe a selected set that is most likely to be realized in finite-temperature systems.

\subsection{Zero field}

In zero field, the simplest solution to the normalization constraints is the coplanar solution introduced in Ref.~\onlinecite{Glittum2023}. It amounts to setting all $\uv$'s and $\vv$'s to zero except $\uv_{01}\perp \vv_{01}$ and $\uv_{23}\perp\vv_{23}$, which are all unit vectors.
Explicitly, it is
\begin{equation}\label{eq:SLP1}
\begin{split}
  \Sv_{\Rv, 0} &= + \uv_{01} \cos{(\Qv_{01} \cdot \Rv)} + \vv_{01} \sin{(\Qv_{01} \cdot \Rv)},\\
  \Sv_{\Rv, 1} &= - \uv_{01} \cos{(\Qv_{01} \cdot \Rv)} - \vv_{01} \sin{(\Qv_{01} \cdot \Rv)},\\
  \Sv_{\Rv, 2} &= + \uv_{23} \cos{(\Qv_{23} \cdot \Rv)} + \vv_{23} \sin{(\Qv_{23} \cdot \Rv)},\\
  \Sv_{\Rv, 3} &= - \uv_{23} \cos{(\Qv_{23} \cdot \Rv)} - \vv_{23} \sin{(\Qv_{23} \cdot \Rv)}.
\end{split}
\end{equation}
This corresponds to sublattice pairing $(01)\&(23)$. Other sublattice pairings ($(02)\&(13)$ and $(03)\&(12)$) are also possible. In this state, the three coplanar spins on a particular plane and sublattice are constructed from a single wave vector $\Qv_{ij}$ (and its negative) and form a 120\degree arrangement which is the same for all planes on the given sublattice. We call this state \SLPONE, where the subscript refers to the number of different momenta $\Qv_{ij}$ on each sublattice ($\pm \Qv_{ij}$ are considered equal in this respect). As a whole, \SLPONE is a $2\Qv$-state, as both $\Qv_{01}$ and $\Qv_{23}$ are present. This \SLPONE state has the characteristic feature of antiparallel spins going along the $\av_0-\av_1$ and $\av_2-\av_3$ directions on sublattices $0,1$ and $2,3$, respectively.
In Ref.~\onlinecite{Glittum2023}, it was found that among the possible relative orientations of the spiral ordering vectors $\uv_{01}$ and $\uv_{23}$ ($\vv_{01}$ and $\vv_{23}$), spin-wave fluctuations selects entropically the \textit{coplanar} state $\uv_{01}=\uv_{23}$ and $\vv_{01}=\vv_{23}$.

The coplanar \SLPONE solution in Eq.~\eqref{eq:SLP1} is a special case of a more general solution where the nonzero $\uv$'s and $\vv$'s are
\begin{align}
\uv_{01} = \uv_{23}, \quad \vv_{01} = \vv_{23} , \quad \vv_{02}=\vv_{12}=\vv_{03}=-\vv_{13}, \label{eq:wsolution1}
\end{align}
with $\uv_{23}$ being a unit vector that is {\em perpendicular} to two other unit vectors $\wv_1$ and $\wv_2$ which parametrize
\begin{align}
\vv_{13}= (\wv_1+\wv_2)/3, \quad \vv_{23}=(2\wv_1-\wv_2)/3. \label{eq:wsolution2}
\end{align}
The presence of the other $\vv$'s in addition to $\vv_{01}$ and $\vv_{23}$ causes the appearance of all six momentum vectors $\Qv_{ij}$ in the spin configuration. Thus, this state is a $6\Qv$-state when $\vv_{13} \neq 0$.
A general property of these states is that spins on sublattices 0 and 1 (2 and 3) with identical trisublattice and plane indices are antiparallel, and spins on sublattices 0 and 2 (1 and 3) with identical trisublattice and plane index are parallel. Additionally, the spins in one of the three planes on the same sublattice are in general different from the spins in the remaining two planes. Thus, the strict up-down alternation along directions $\av_0-\av_1$ and $\av_2-\av_3$ does in general no longer hold.

The special case $\wv_1=-\wv_2$ implies $\vv_{13}=0$ and corresponds to the coplanar \SLPONE solution above, where the three planes on each sublattice are equal. For  $\wv_1=\wv_2$, we find a coplanar state that differs slightly from \SLPONE. In this state, two spins on one of the triangular planes on every sublattice are permuted, so that the chirality of the 120\degree arrangement in that plane is opposite to the remaining two planes.
%and we find that for this state $O_{\rm SLP}=1/3$.

For noncollinear arrangements of $\wv_1$ and $\wv_2$, one gets a noncoplanar spin state. Because of the noncoplanarity, we expect this state to have smaller entropy than the coplanar \SLPONE state,  and less likely to be realized at finite temperatures.

\subsection{Finite field}

In a finite magnetic field, the general zero-field solution Eqs.~(\ref{eq:wsolution1},\ref{eq:wsolution2}) gives unit length spins if $\uv_{23},\wv_1$ and $\wv_2$ are modified to be vectors of length $\sqrt{1-m^2}$ perpendicular ($\perp$) to the field. This implies a configuration where the $\perp$-components of the spins with identical trisublattice and plane indices are antiparallel on sublattices 0 and 1 (2 and 3) and parallel on sublattices 0 and 2 (1 and 3). The components parallel ($\para$) to the field of all the spins are identical and equal to the magnetization per spin $m$.
However, since the vectors $\uv_{23}$, $\wv_1$ and $\wv_2$ lie in the plane perpendicular to the magnetic field and $\uv_{23}$ must be perpendicular to both $\wv_1$ and $\wv_2$, there are only two possibilities for $\wv_1$ and $\wv_2$ given $\uv_{23}$. In the first case, the $\wv$'s are antiparallel, giving a canted version of \SLPONE. This state is essentially \SLPONE within the $\perp$-plane, but with all spins having the same finite $\para$-component (see Fig.~\ref{fig:SLP1}).
%In this canted \SLPONE state $O_{\rm SLP}=1-\mv^2$.
The other case, parallel $\wv$'s, gives similarly a canted version of the zero-field state where one plane of spins has an opposite chirality in-plane 120\degree arrangement than in the other two planes.
%and $O_{\rm SLP}=(1-\mv^2)/3$.

The previously discussed noncoplanar zero-field solutions have no analog in a magnetic field. However, there are other classes of solutions which resemble the previous ones in that any spin on a given sublattice is ``antiparallel'' to one spin on two of the other sublattices (in a field we use the term ``antiparallel'' for two spins having opposite $\perp$-components and equal $\para$-components) and parallel to a spin on the remaining sublattice.
Such solutions are more likely to have high entropy due to the high number of (anti)parallel spins.
We have found one particularly interesting example of this, where spins that are ``antiparallel'' do no longer have equal plane indices, as was the case for the solutions discussed previously. Instead, the spins in planes $(0, 1, 2)$ on sublattice 0 (2) are ``antiparallel'' to the spins of equal trisublattice index in planes $(0, 2, 1)$ on sublattice 1 (3), which leads to
\begin{align*}
u_{12}^{\perp} = -u_{03}^{\perp}, \;\; &u_{12}^{\para} = +u_{03}^{\para}, \\
v_{12}^{\perp} = +v_{03}^{\perp}, \;\; &v_{12}^{\para} = -v_{03}^{\para}, \\
u_{13}^{\perp} = -u_{02}^{\perp}, \;\; &u_{13}^{\para} = +u_{02}^{\para}, \\
v_{13}^{\perp} = -v_{02}^{\perp}, \;\; &v_{13}^{\para} = +v_{02}^{\para},  \\
u_{01}^\para = 0, \;\; v_{01}^\para = 0, \;\; &u_{23}^\para = 0, \;\; v_{23}^\para = 0.
\end{align*}
Further, the spins on sublattices 0 (1) and 2 (3) are parallel for equal plane and trisublattice indices, implying
\begin{align*}
\uv_{02} &= 0,\\
u_{03}^\para &= 0, \;\;\;\;\; v_{03}^\para = 0,\\
\uv_{23} &= \uv_{01},\;\; \vv_{23} = \vv_{01}.
\end{align*}
This leaves us with only two vectors with finite $\para$-components: $\vv_{02}$ and $\vv_{13}$. The remaining $\uv$'s and $\vv$'s all lie in the $\perp$-plane.

\begin{figure*}
\begin{minipage}{\columnwidth}
\includegraphics[width=\columnwidth]{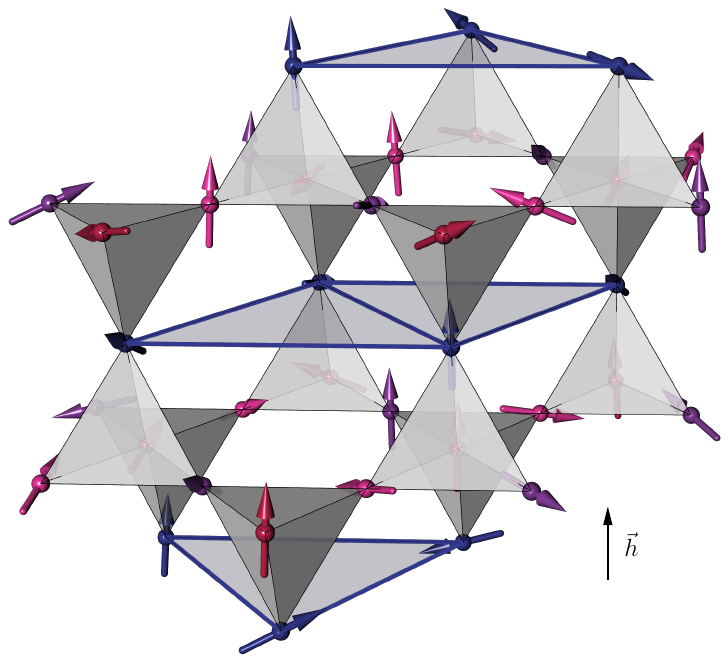}
\end{minipage}
~
\begin{minipage}{\columnwidth}
\begin{center}
\includegraphics[width=\columnwidth]{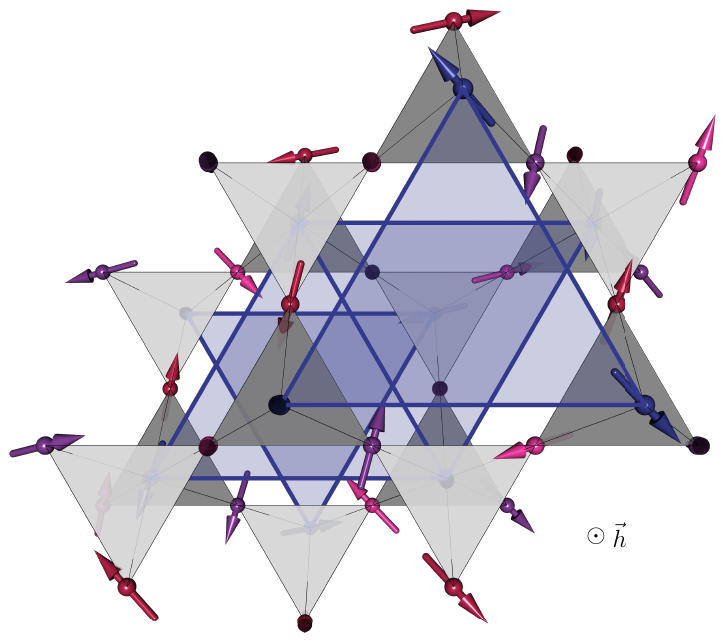}
\end{center}
\end{minipage}
\caption{\markup{The \SLPTHREEM state in magnetic field $h=4$. Left: The spins have three different $\para$-components, and for each sublattice there are three different triangular plane spin configurations that repeats itself with period three.  Right: The same configuration, but rotated such that the magnetic field points out of the paper. This illustrates how the spins in every triangle sums to zero in the $\para$-plane. We also see how some spins are collinear in this plane.}}
\label{fig:SLP3}
\end{figure*}

Even with these conditions, there are many solutions. However, solutions different from those already classified have $\vv_{02}^\perp = 0$.
The  normalization problem then reduces to nine equations (normalization of the nine spins on sublattice zero) with eleven unknowns. We solve these nine equations, leaving the direction of $\uv_{01}$ and the angle $\alpha - \pi/2$ between $\uv_{01}$ and $\uv_{03}$ as free parameters. We find that $\uv_{01}$ ($\uv_{03}$) and $\vv_{01}$ ($\vv_{03}$) must be orthogonal vectors of length $\sqrt{a + b}$ ($\sqrt{a - b}$), where
\begin{align*}
a &= \frac{1}{2}(1-m^2-8m^2\tan^2 \alpha),\\
b &= \frac{1}{2}\sqrt{(1-m^2)^2-16m^2(1+3m^2)\tan^2 \alpha}.
\end{align*}
Additionally, the directions of $\vv_{01}$ and $\vv_{03}$ must be chosen to satisfy
\begin{align*}
\uv_{01} \cdot \vv_{03} &= \uv_{03} \cdot \vv_{01}.
\end{align*}
The nonzero component of  $\vv_{02}$ is $v_{02}^\para = -4 |\vec{m}| \tan \alpha$.
Lastly, the angle $\alpha$ is restricted by
\begin{equation}
\alpha \in [0, \alpha_{\rm max}] \cup [ \pi - \alpha_{\rm max},  \pi],
\end{equation}
where
\be
\tan\alpha_{\rm max} = \f{1-m^2}{4|m|\sqrt{(1+3m^2)}}. \label{eq:alphamax}
\ee
For $\alpha=\{0,\pi\}$, this solution is the canted \SLPONE state with $\uv_{01}^2 = 1-m^2$ and $\uv_{03}^2 = \vv_{02}^2 = 0$. As $\alpha$ is increased (decreased) from $0$ ($\pi$), the canted \SLPONE state continuously evolves into a state where each sublattice has three $\Qv$s; \SLPTHREEALPHA (with six $\Qv$s in total).
In this state, there is a different order in each of the three triangular planes on a given sublattice. An explicit parametrization of this \SLPTHREEALPHA state is shown in Appendix~\ref{app:explicitSLP3}. The value $\alpha_{\rm max}$ gives the ``maximal'' \SLPTHREEALPHA state, where $|\uv_{01}| = |\uv_{03}|$. We will refer to this as the \SLPTHREEM state (see Fig.~\ref{fig:SLP3}).

In the \SLPTHREEALPHA states, the spin components along the magnetic field take three values: $\{m,m(1 \pm 2\sqrt{3} \tan{\alpha})\}$. This is a consequence of only $\vv_{02}$ and $\vv_{13}$ having a finite $\para$-component.  For $\alpha=\{0,\pi\}$, all spin $\para$-components become equal to $m$ (canted \SLPONE). The difference in the three values along the field increases as $\alpha$ approaches $\alpha_{\rm max}$.

Taking the limit $m\to 0$, we note that \SLPTHREEALPHA approaches the coplanar single-$\Qv$ \SLPONE state for all $\alpha \neq \alpha_{\rm max}$, while for $\alpha = \alpha_{\rm max}$, we instead get a noncoplanar zero-field analog of the \SLPTHREEM state, different from the previously discussed zero-field solutions.

We note that other solutions similar to \SLPTHREEALPHA can be found by transforming $\Rv$ according to the lattice symmetries. This will lead to solutions with other pairings of ``antiparallel'' and parallel planes and trisublattice indices, depending on the transformation.

\section{Spin-wave entropy \label{sec:spinwaveentropy}}
For a given value of the magnetic field, all states described by the ansatz in Eq.~(\ref{eq:general ansatz}) have equal energies. To determine which of these are favored at finite temperatures, we compute their entropy using linear spin-wave theory. We assign each particular ground state to a periodic fcc system with the unit cell of $M=3^3 \times 4 =108$ spins. The spin configurations define site-dependent rotated frames, in terms of which the spins are in a ferromagnetic configuration. We then use the Holstein-Primakoff~\cite{HolsteinPrimakoff1940} transformation to describe the deviations from the ferromagnetic configuration as bosons. The resulting boson Hamiltonian is expanded to quadratic order, with vanishing linear terms. This Hamiltonian is then diagonalized numerically to obtain the spin-wave frequencies $\omega_{\qv,i}$, using a Bogoliubov transformation involving $2M$ bosons~\cite{Colpa1978}.  \add{We regularize the zero modes by adding a small value $10^{-10}$ to the diagonal of the Hamiltonian matrix.} For further details, see Appendix~\ref{app:spinwaves}.

The entropy per site from these linear spin-wave fluctuations can then be computed as
 \be
     S/N = -\f{1}{l^3 M} \sum_{\qv} \sum_{i=1}^{M} \ln{\omega_{\qv,i}},
\ee
where the \markup{sum over $\qv$} goes over the first Brillouin zone of the lattice with $l^3$ $M$-site unit cells. In order to find the entropy for $l \to \infty$, we compute the entropy for sizes $l=\{6,9,12,15,18\}$ and extrapolate $S/N$ versus $1/l^3$ to infinite size using a quadratic polynomial.
\add{We use the parameter values $J_1=1$, $J_{3b}=0.2$.}

\markup{In zero magnetic field, the ground states are parametrized by three unit vectors: $\uv_{23},\wv_1$ and $\wv_2$, where $\uv_{23}$ is perpendicular to $\wv_1$ and $\wv_2$. We label the angle between $\wv_1$ and $\wv_2$ as $\theta$. Then, computing the entropy for these solutions, we get the results shown in Fig.~\ref{fig:h0entropy}. The maximal entropy occurs for $\theta=\pi$, which corresponds to the coplanar \SLPONE solution.

For the zero-field \SLPTHREEM state, we find eight gapless modes, compared to 14 for the coplanar \SLPONE state. Thus, we conclude that the coplanar \SLPONE solution also has higher entropy than the zero-field \SLPTHREEM state.

\begin{figure}
\begin{center}
\includegraphics[width=\columnwidth]{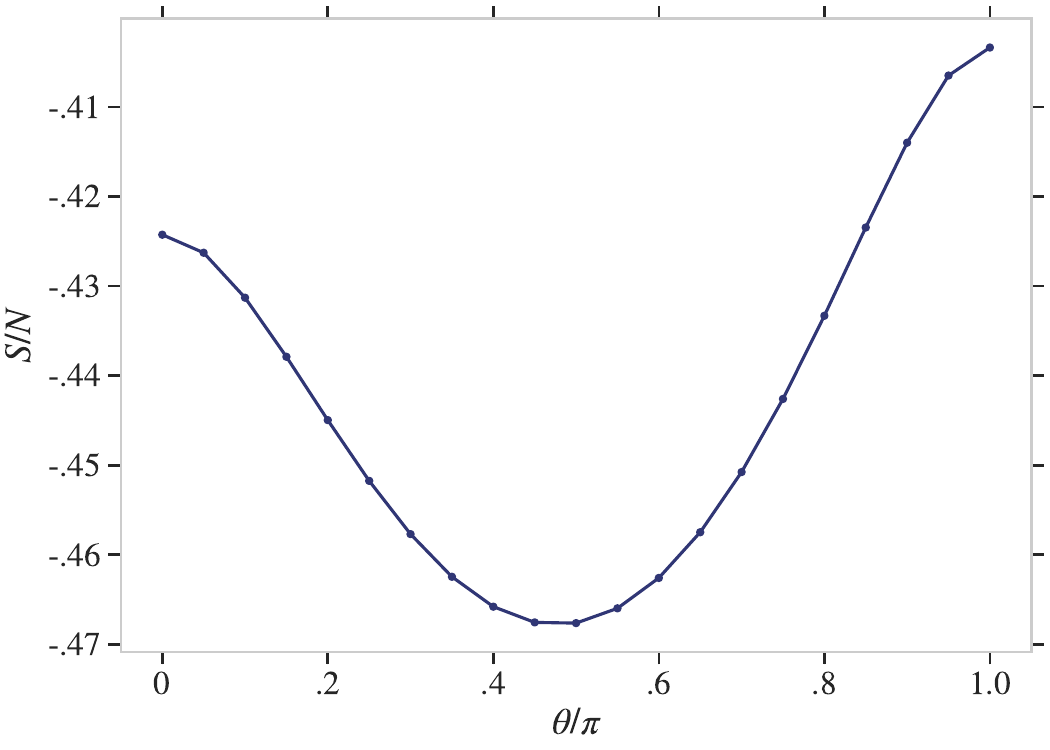}
\caption{Entropy for the $h=0$ solution characterized by the angle $\theta$ between the two vectors $\wv_1$ and $\wv_2$. \label{fig:h0entropy}}
\end{center}
\end{figure}
}

\begin{figure}
\begin{center}
\includegraphics[width=\columnwidth]{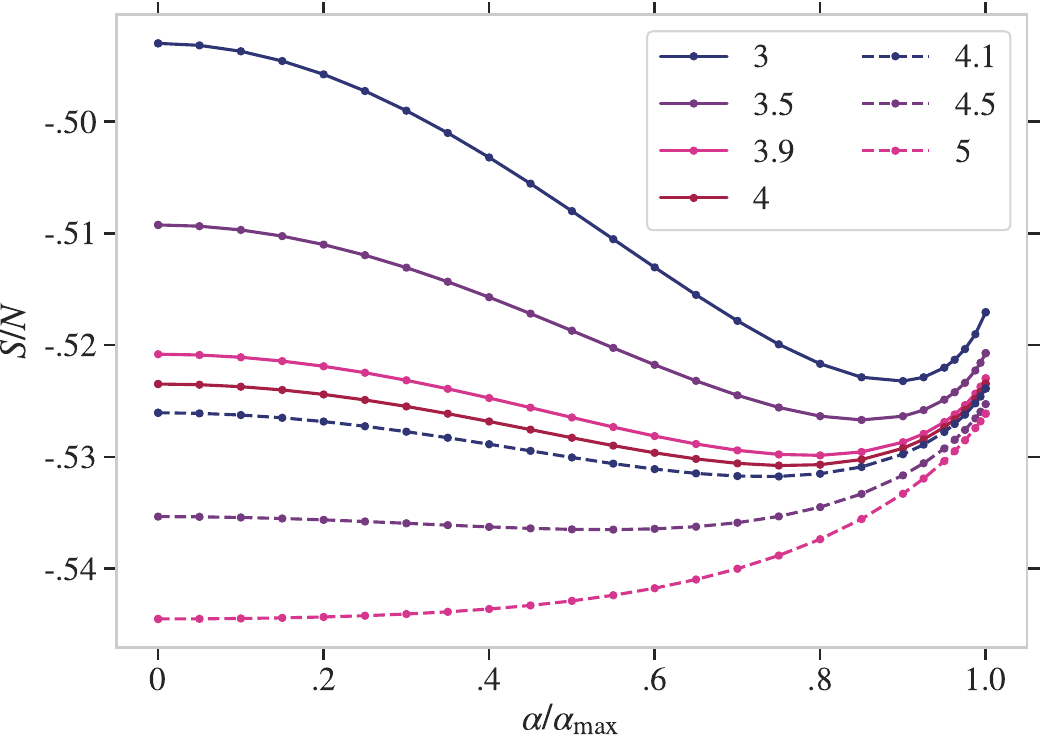}
\caption{Entropy per spin versus $\alpha$ in the \SLPTHREEM state for different values of the magnetic field (legend). \delete{$J_1=1$, $J_{3b}=0.2$.}}
\label{fig:entropy}
\end{center}
\end{figure}

\begin{figure}
\begin{center}
\includegraphics[width=\columnwidth]{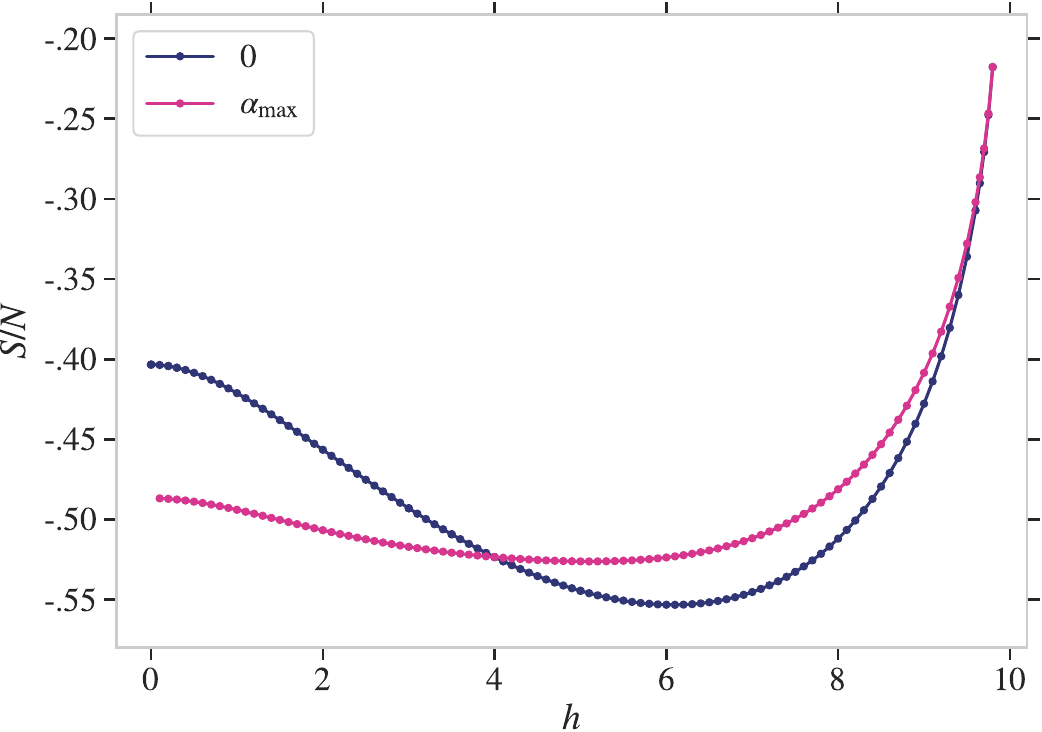}
\caption{Entropy versus magnetic field for $\alpha=0$ and $\alpha=\alpha_{\rm max}$. \delete{$J_1=1$, $J_{3b}=0.2$.}}
\label{fig:entropyvsfield}
\end{center}
\end{figure}

For finite fields, \delete{we compute} the entropy per site \delete{(for $l \to \infty$)} for \SLPTHREEALPHA as a function of $\alpha$, \delete{as}\add{is} shown in Fig.~\ref{fig:entropy}\delete{ for $J_1=1$ and $J_{3b}=0.2$}. The Bogoliubov diagonalizations for these configurations yield four gapless modes. It is seen that for low fields ($h<4$), the entropy is maximal for $\alpha=0$, corresponding to the canted \SLPONE state. For higher fields ($h>4$), the entropy is instead maximal for $\alpha = \alpha_{\rm max}$, corresponding to \SLPTHREEM. Figure~\ref{fig:entropy} also shows that the change between these two values ($0$ and $\alpha_{\rm max}$) at $h=4$ is discontinuous.
To better see that the change happens at $h=4$, we compare the values of the entropy per site versus magnetic field for the two $\alpha$ values in Fig.~\ref{fig:entropyvsfield}.

\section{Order parameters \label{sec:orderparameters}}
To distinguish the \SLPONE state from the \SLPTHREE states in the subsequent numerics, we construct \delete{three}\add{four} order parameters. For the first order parameter, we make use of the fact that
the \SLPONE state can be viewed as two sets of crossing parallel chains where the spins on each chain are ordered in an ``antiparallel'' way. In an \SLPONE state with $(01)\&(23)$ pairing, one set of chains runs along the direction $\av_1-\av_0$ involving alternate sublattice 0 and 1 sites. The other set runs along the direction $\av_3-\av_2$ and involves sites on the sublattices 2 and 3 (see Fig.~\ref{fig:SLP1}). For one chain involving sublattices $i$ and $j$, we define the chain staggered order parameter as
\be
  \vec{C}_{\Rv_0, ij} \equiv  \f{1}{2L} \sum_{n} \left( \Sv_{\Rv_0 + n(\av_j-\av_i), i} - \Sv_{\Rv_0 + n(\av_j-\av_i), j} \right),
\ee
where $\Rv_0$ is an arbitrary up tetrahedron on the chain. \add{$L$ is the linear size of the system such that the total number of spins is $N=4L^3$.} 
We take the average length of this over all chains, which is achieved by taking the average over all up tetrahedra in the lattice
\be
   C^2_{ij} \equiv \f{1}{L^3} \sum_{\Rv_0}  \vec{C}_{\Rv_0, ij} \cdot \vec{C}_{\Rv_0, ij}.
\ee
From this, we construct the three-component complex SLP order parameter, $O_{\rm SLP}$, where each component corresponds to a particular way of pairing the sublattices,
\be
O_{\rm SLP} = C_{01} C_{23} + e^{i 2\pi/3} C_{02} C_{13} +e^{i 4\pi/3} C_{03} C_{12}.
\ee
In the MC simulations, we measure all the $C_{ij}$s and compute the modulus $|O_{\rm SLP}|$. A nonzero value will be taken as an indicator of an SLP phase.
In Appendix~\ref{app:order parameter} it is shown that for a state respecting the SLP ansatz in Eq.~\eqref{eq:general ansatz},
\be
   C^2_{ij} = \f{1}{2} \left( \uv_{ij}^2  + \vv_{ij}^2 \right)
\ee
and
 \begin{align}
 |O_{\rm SLP}|  &= \sqrt{ \left(1 - \mv^2 \right)^2 - 3 \left( C^2_{01} C^2_{02} + C^2_{01} C^2_{03} +C^2_{02} C^2_{03} \right)}.
 \end{align}
The maximum value occurs for the canted \SLPONE state, where only one of $C_{01}$, $C_{02}$ and $C_{03}$ is nonzero, giving
\begin{align}
|O_{{\rm SLP},1}|  &= 1 - m^2.
\end{align}
In the \SLPTHREEM state, we get
\begin{align}
  |O_{{\rm SLP},M}|  &= \left| \f{(1 - m^2)(1-9m^2)}{4(1+3m^2)} \right|. \label{OSLPinSLP3}
\end{align}

Another property of the canted \SLPONE state is that all spin $\para$-components are equal. To detect deviations from this, we calculate the spatial fluctuations of the magnetization in the field-direction
\be
\kappa \equiv \f{1}{N} \sum_{\rv} \left( S^\para_{\rv} -m \right)^2.
\ee
For the canted \SLPONE state; $\kappa =0$, while for the \SLPTHREEM state
\be
\kappa_M = \f{(1-m^2)^2}{2(1+3m^2)}. \label{KappainSLP3}
\ee

\delete{Lastly,} \SLPONE and \delete{\SLPTHREE}\add{\SLPTHREEALPHA} can also be distinguished by only \delete{\SLPTHREE}\add{\SLPTHREEALPHA} having a nonvanishing tetrahedral scalar chirality~\cite{Berg1981,Aoyama2021}
\be
 O_\chi = \f{4}{N} \sum_{\Rv} \chi(\Rv),
\ee
where $\chi(\Rv)$ is the sum of the chiralities for the four faces of the up tetrahedron at $\Rv$. The chirality of a face is computed such that its three spins are ordered in a counter-clockwise direction w.r.t. the outward face normal:
\begin{align}
\begin{split}
\chi(\Rv) &= \Sv_{\Rv,1} \cdot ( \Sv_{\Rv,2} \times \Sv_{\Rv,3} )+  \Sv_{\Rv,3} \cdot ( \Sv_{\Rv,0} \times \Sv_{\Rv,1} ) \\
                  \quad &+ \Sv_{\Rv,0} \cdot ( \Sv_{\Rv,3} \times \Sv_{\Rv,2} )+  \Sv_{\Rv,2} \cdot ( \Sv_{\Rv,1} \times \Sv_{\Rv,0} ).
\end{split}
\end{align}
For the \SLPTHREEALPHA state, $O_\chi=  2 \left( 4m \tan{\alpha} \right)^3$, which means that it is zero for the canted \SLPONE state ($\alpha=0$) and
\be
O_{\chi, M} =   2 \left( \f{1- m^2}{\sqrt{1+ 3m^2}} \right)^3
\ee
for \SLPTHREEM.

\add{A way to detect ordered multi-$\Qv$ states is to measure the product of the static structure factors at the different ordering wave vectors.
The static structure factor
\be
      S^{\mu \nu}(\qv) \equiv \f{1}{2} \left( S^{\mu}_{-\qv} S^{\nu}_{\qv} + S^{\nu}_{-\qv} S^{\mu}_{\qv} \right) \label{def_staticstructurefactor}
\ee
is defined in terms of the Fourier transform $S^{\mu}_{\qv}$ of the spin state. We define the (isotropic) $3 \Qv$ (on each sublattice) order parameter which detects simultaneous ordering at $\Qv_{01}$, $\Qv_{02}$ and $\Qv_{03}$ in the 111 Brillouin zone as
\be
    O_{3\Qv} = \f{1}{N^3} \left( S^{\lambda \lambda}(\Qv_{01}+\Gv^\prime) S^{\mu \mu}(\Qv_{02}+\Gv^\prime) S^{\nu \nu}(\Qv_{03}+\Gv^\prime) \right)^{1/2},
\ee
where repeated Greek indices are summed over, and $\Gv^{\prime} \equiv \bv_1+\bv_2+\bv_3$.  $O_{3\Qv}$ is zero in the \SLPONE phase, and nonzero in \SLPTHREE phases; see Sec.~\ref{sec:staticsf}.
}     

%Since the \SLPTHREE state is a multi-$\qv$ state, and $\chi$ is a product of three spins on different sublattices, $\chi(\qv)$ will be nonzero in the \SLPTHREE state for many values of $\qv$ corresponding to sums and differences of the $\Qv_{ij}$ vectors. We find that for the maximal $\alpha$-value allowed for a given magnetization the maximal value of $\chi_{\qv} \chi_{-\qv}$ occurs for $\qv = b_1 + b_2$ and symmetry related values which corresponds to summing all chiralities for the up tetrahedra and subtracting the sum on the down tetrahedra. This motivates the choice of our order parameter for the three plane multi-q phase as
%\be
%   O_{\chi} = \f{1}{2L^3} \sum_{\Rv} \left( \chi_{u,ijk}(\Rv)- \chi_{d,ijk}(\Rv) \right)
%\ee
%where $u$ ($d$) refers to the up (down) tetrahedron in unit cell $\Rv$.

\section{Numerical Results \label{sec:results}}
To get an independent confirmation of our analytical results, we have performed classical MC simulations on systems with an fcc lattice of tetrahedra with periodic boundary conditions.
The simulations were carried out using \add{multiple MC update types. In all our simulations one MC step consisted of $N$} single-spin Metropolis moves~\cite{Metropolis1953},
where \markup{for each spin the proposed new random direction is restricted to lie in a cone about the old spin~\cite{LandauBinder2009,Hinzke1999,Alzate-Cardona2019}\delete{ random vector in spin space~\cite{LandauBinder2009}. As the acceptance ratio of these Metropolis updates were only a few percent at the lowest temperatures, we included also other types of MC updates:},}
\add{$N$ o}ver-relaxation moves~\cite{LandauBinder2009} where
\add{a random}\delete{the} spin is rotated an arbitrary angle about the effective magnetic field seen on its site,
a \add{number of} Wolff-cluster updates~\cite{Wolff1989} where clusters are formed of spins that will be mirrored on an arbitrary plane that includes the magnetic field direction (so that $S^\para$ does not change),
and $3N$ single-spin Metropolis moves where the cartesian spin components,
\add{$S^x,S^y$ and $S^z$, of a single spin is attempted}\delete{ flipped}\add{ reversed.}\delete{, and a single-spin Metropolis move where the proposed change is restricted to lie in a cone about the old spin.} In some of the simulations, \add{in addition to the above MC updates,} we also implemented the parallel tempering setup~\cite{SwendsenWang1986} with a feedback-optimized scheme~\cite{Katzgraber2006} to select a set of temperatures that cause as many round-trips as possible for each replica. The parallel tempering simulations were run with a fixed number of typically 20-50 replicas spanning a fixed temperature interval, with more replicas if the system is large in order to increase the replica swapping rate.

\begin{figure}
\begin{center}
\includegraphics[width=\columnwidth]{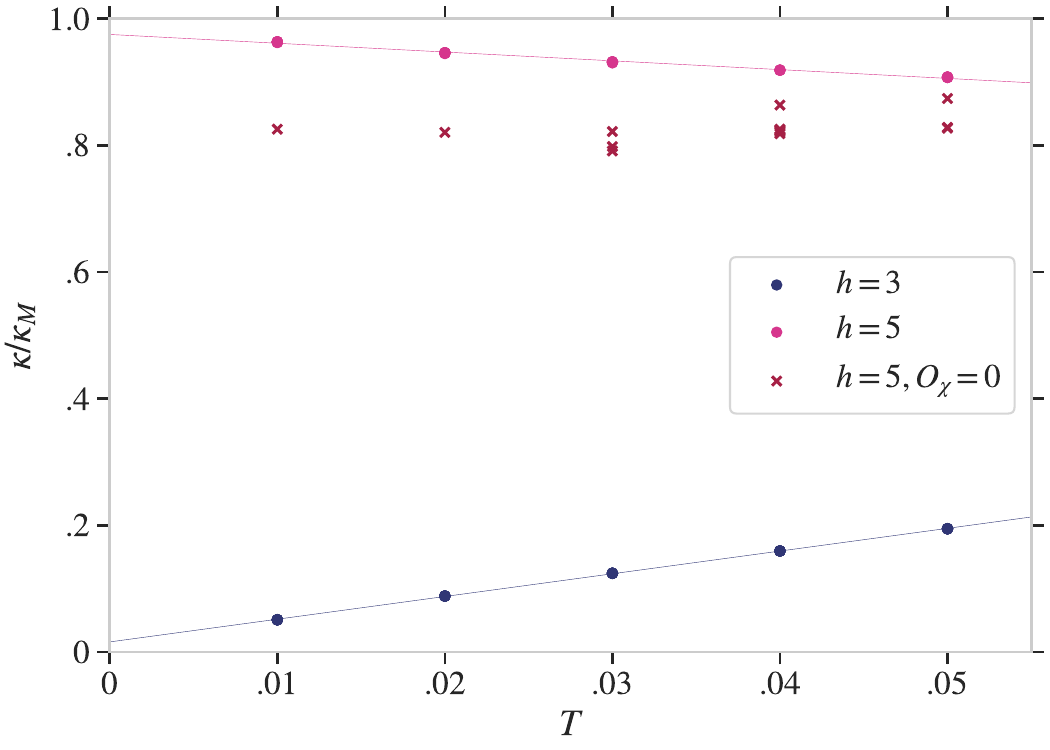}
\caption{Scaled magnetization fluctuations $\kappa$ at low temperatures for magnetic fields $h=3$ and $5$. $L=6$. Many points are on top of each other as we show the results for ten independent runs for each temperature and field value.}
\label{fig:lowT}
\end{center}
\end{figure}

In the following we will show results for the parameter values $J_1=1$ and $J_{3b}=0.2$. \add{The MC results shown in the figures (except Fig.~\ref{fig:lowT}) are mean values of 10-20 statistically independent bins, and the error bars mark the standard deviation of the mean. Where not visible, the error bars are smaller than the symbol size.}

All our low-temperature numerical simulations show results consistent with the magnetization relation Eq.~\eqref{eq:magnetization} and ground state energy Eq.~\eqref{eq:groundstateenergy}, confirming the form of the Hamiltonian in Eq.~(\ref{eq:Hamiltonian2}).

\begin{figure}[]
\includegraphics[width=\columnwidth]{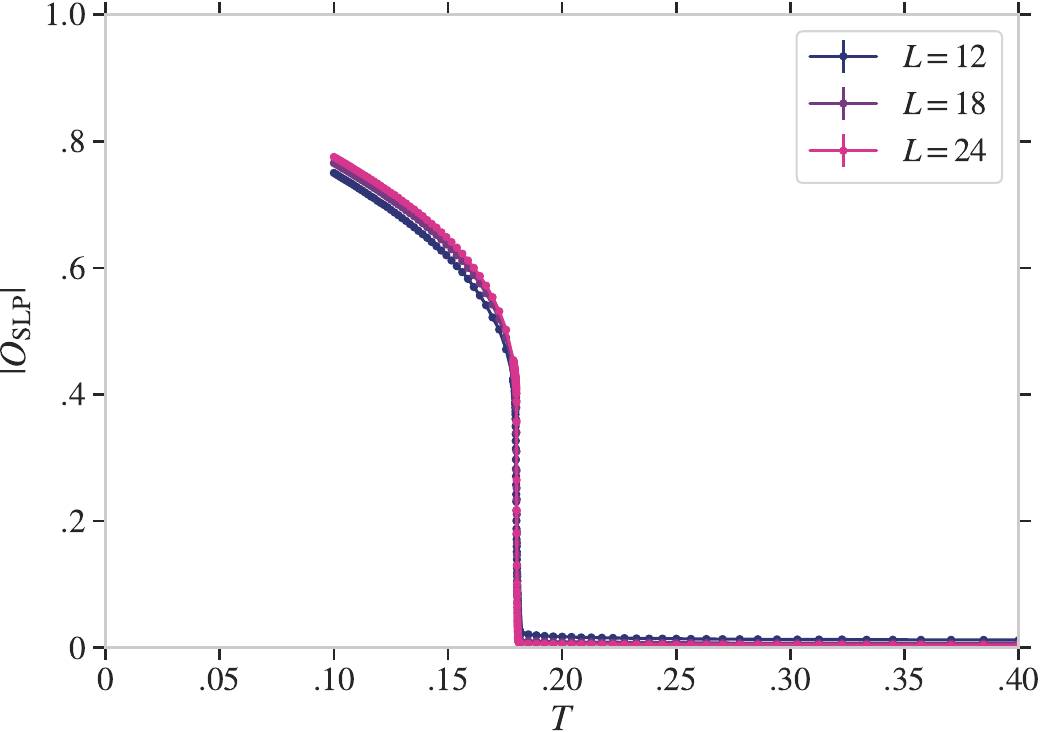}
\caption{Parallel tempering Monte Carlo results for the SLP order parameter versus temperature in zero magnetic field. \label{SLP120}}
\end{figure}

In Fig.~\ref{SLP120}, we show parallel tempering MC results for $|O_{\rm SLP}|$ versus temperature $T$ in zero magnetic field for three different linear system sizes $L$ ($N=4L^3$).  The SLP order parameter shows a
steep increase at $T_c=0.180$, which is $8\%$ lower than estimated using nematic bond theory~\cite{Glittum2023}. %NBT: 0.195
The system size dependence of the critical temperature is weak, and the rise of the order parameter resembles that of a discontinuous phase transition, in accordance with the nematic bond theory results.  When extrapolating to zero temperature, the SLP order parameter appears to reach the value 1, consistent with the low-temperature phase being the \SLPONE state (with thermal fluctuations).

\begin{figure}[]
\includegraphics[width=\columnwidth]{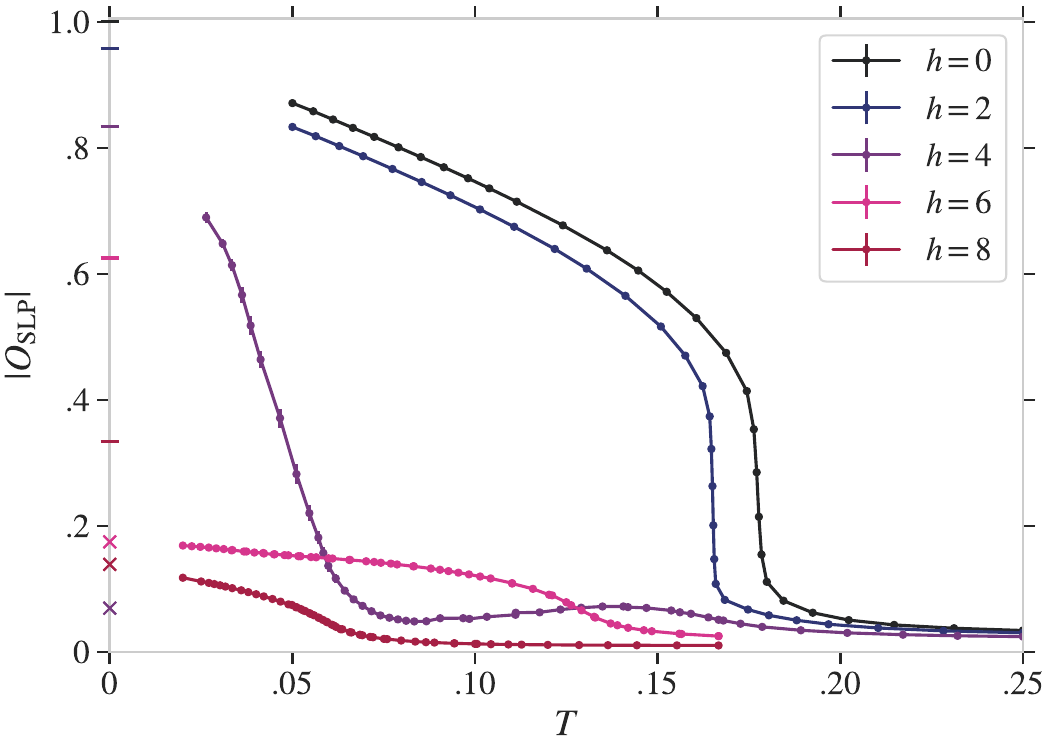}
\caption{Parallel tempering Monte Carlo results for the SLP order parameter versus temperature for different values of the magnetic field indicated in the legend. $L=6$.
The short horizontal colored bars on the ordinate axis indicate the \SLPONE values of the SLP order parameter; $1-m^2$. The colored crosses on the ordinate axis indicate the \SLPTHREEM values of the SLP order parameter from Eq.~\eqref{OSLPinSLP3}.
\label{PT6}}
\end{figure}

When a magnetic field is added, the MC simulations are in general more difficult to equilibrate\delete{, thus we were forced to use smaller system sizes}. Using a system size of $L=6$, we show in Fig.~\ref{PT6} the SLP order parameter versus temperature for different values of the magnetic field.
The sharp rise of the SLP order parameter that was seen for zero field is still present for fields $h \lesssim 4.5$, and it appears to extrapolate to a value close to $1-m^2$ as $T\to 0$, shown as colored horizontal bars on the ordinate axis. This is consistent with the low-temperature phase being the canted \SLPONE state. The associated critical temperature decreases as the magnetic field is increased. 

\begin{figure}[ht]
\includegraphics[width=\columnwidth]{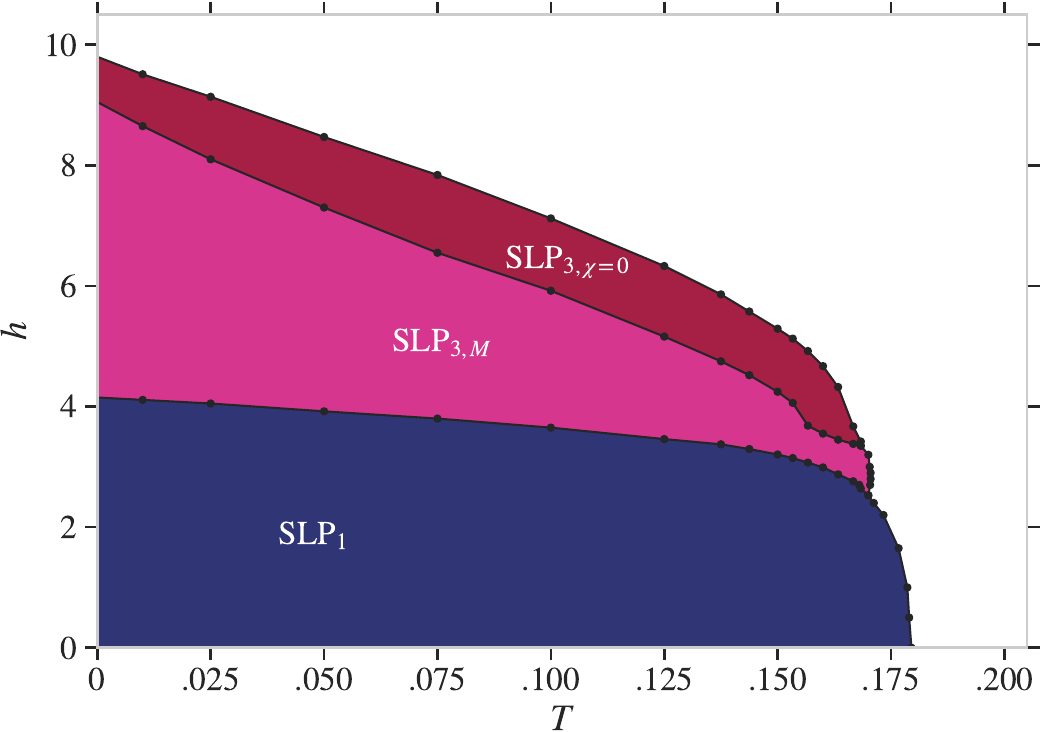}
\caption{Phase diagram for $J_1=1$ and $J_{3b}=0.2$. $L = 24$. \label{fig:phasediagram}}
\end{figure}

We also observe from Fig.~\ref{PT6} that for small fields, $h \lesssim 2\add{.5}$, the system goes directly from the high-temperature disordered phase into the canted \SLPONE phase at low temperatures.
In contrast, for $2\add{.5} \lesssim h \lesssim 4.5$ coming from high temperatures, there is first a slight increase in the SLP order parameter before it slightly decreases, and then increases sharply at a lower temperature (see the $h=4$ curve in Fig.~\ref{PT6}). For $h \gtrsim 4.5$ we find only the slight increase in the SLP order parameter, and no dramatic upturn down to the lowest temperatures studied ($T>0.02$). Instead, the SLP order parameter extrapolates to a value close to that for the \SLPTHREEM state when $T$ goes to zero, shown as crosses on the ordinate axis. We thus interpret the slight increase of the SLP order parameter as entering the \SLPTHREEM phase and the more pronounced increase as entering the \SLPONE phase. \delete{To preliminary summarize these data, we draw in Fig.~\ref{fig:phasediagram} an outline of the phase diagram, which will be further substantiated below.}

To \markup{\delete{verify the \SLPTHREEM part of the phase diagram} investigate the stability of the canted \SLPONE state in a magnetic field} at low temperatures, we have carried out long standard MC simulations without parallel tempering, all started from the canted \SLPONE state. They were run for $10^9$ MC steps, each at a fixed temperature. In Fig.~\ref{fig:lowT}, we show the spatial magnetization fluctuations $\kappa$ (scaled by the expected value $\kappa_M$ in the \SLPTHREEM state) for ten independent simulations at each temperature and field value. It is apparent that for $h=5$, all simulations have escaped the initial \SLPONE state. Most of them (circles) get a value for $\kappa$ which comes very close to the \SLPTHREEM value when $T \to 0$. There are also runs with slightly smaller values of $\kappa$ (crosses), that are further distinguished from the others (circles) by having at least an order of magnitude smaller chirality $|O_{\chi}|$.

Taking the final \add{$h=5$ }spin configurations of these runs, and exposing them to an energy relaxation procedure, we relax the spin configurations to the corresponding ground state (relaxed energies differ from Eq.~\eqref{eq:groundstateenergy} by at most $10^{-12}$).
%, thus reducing the effect of thermal fluctuations.
We find that the spin configurations corresponding to higher $\kappa$ relax to \SLPTHREEM, while the ones with smaller $\kappa$ attain a \add{similar, but} more complicated noncoplanar \markup{$6\Qv$} zero chirality configuration\add{,\SLPTHREECHIZERO}, that satisfies the ansatz Eq.~\eqref{eq:general ansatz}, \add{and has ``antiparallel'' spin pairs}, but which is still not among the solutions outlined in Sec.~\ref{sec:solutions}.  While the chirality on individual up tetrahedra are generally nonzero in this state, their sum is zero. We have calculated the entropy per site of \delete{this}\add{such} zero chirality state\add{s} \delete{to be $-0.53056$, which is lower than the value $-0.526129$ obtained for \SLPTHREEM ($h=5$).}\add{ for a range of field values, and find that it is lower than that of \SLPTHREEM up to very high fields, $h \simeq 9.0$.}  Thus, we expect that finite temperature entropy selection will eventually prefer the \SLPTHREEM state for even longer MC runs \add{as long as $h \lesssim 9.0$. For higher magnetic fields the linear spin-wave entropies of the zero chirality state and the \SLPTHREEM state are almost equal.} For $h=3$, $\kappa$ goes to zero as $T \to 0$ for all runs. This is consistent with \SLPONE, which is also what the energy relaxation procedure on these states shows.

\begin{figure}
\begin{center}
\includegraphics[width=\columnwidth]{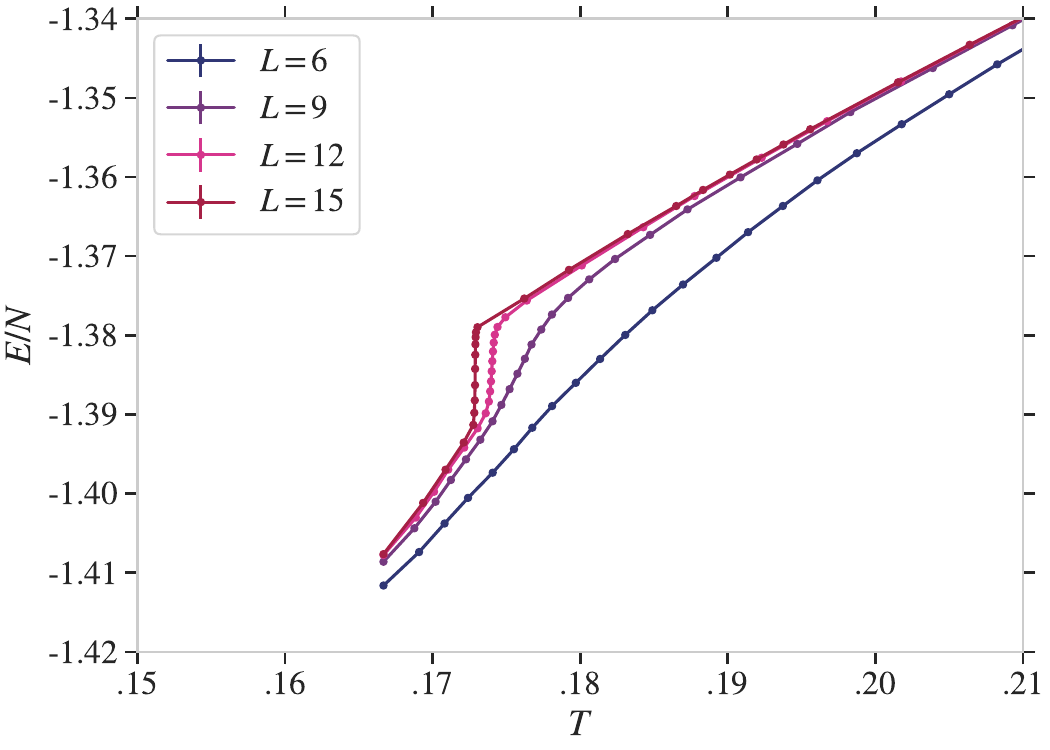}
\caption{Parallel tempering Monte Carlo results for the energy per site versus temperature around the \delete{high-temperature }phase transition for $h=2.5$ for different system sizes.}
\label{fig:pthump}
\end{center}
\end{figure}

\begin{figure}
\begin{center}
\includegraphics[width=\columnwidth]{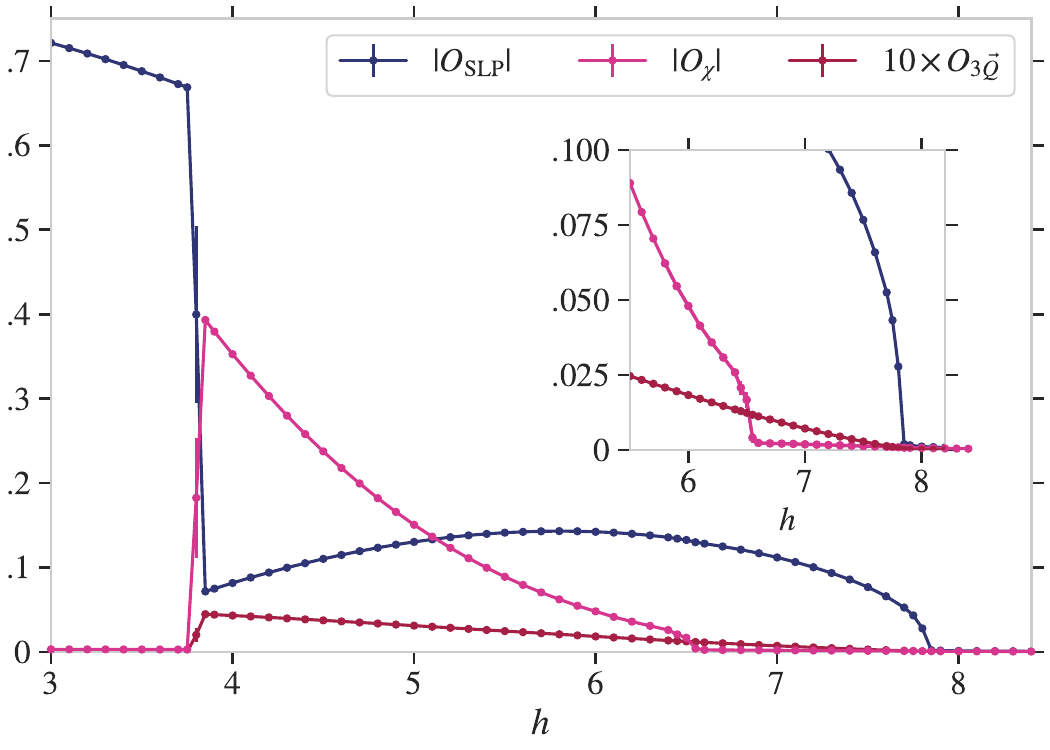}
\caption{\add{The SLP order parameter $|O_{\rm SLP}|$, the chirality order parameter $|O_{\chi}|$ and the $3\Qv$ order parameter $O_{3\Qv}$ (multiplied by a factor 10) versus magnetic field $h$ for fixed $T=0.075$ obtained from mixed phase Monte Carlo simulations. $L=24$. The inset shows a blowup of the high field region.}}
\label{fig:hscan}
\end{center}
\end{figure}

\add{The phase transition into the \SLPONE state at zero magnetic field is discontinuous. This is also the case for small magnetic fields as shown in Fig.~\ref{fig:pthump}, where we show the energy per site at $h=2.5$ versus temperature for different system sizes obtained using the parallel tempering method. At $T \simeq 0.175$ the curves develop with system size into a ``knee'' discontinuity, indicating a first order phase transition.
  In contrast to the zero field case, significant finite size effects are seen at $h=2.5$ and it is therefore necessary to carry out simulations on larger lattices to map out the phase diagram. To do so, we employ the mixed phase technique~\cite{Creutz1979} where fixed temperature MC simulations are started in a configuration where half of a large $L=24$ system is in the canted \SLPONE state and the other half in the \SLPTHREEM state. The runs are relatively short, $\sim 10^4$ MC steps, and the error bars are gotten from repeating the simulations five times with different random seeds. Figure~\ref{fig:hscan} shows the SLP order parameter, the chirality order parameter, and the $3\Qv$ order parameter as functions of magnetic field for a fixed low temperature $(T=0.075)$. For $h < 3.8$ the SLP order parameter is large, and both the chirality and the $3\Qv$ order parameters are zero, consistent with the \SLPONE phase. For $h=3.8$ the $3\Qv$ order parameter makes a jump and then decreases with increasing field and vanishes at $h=7.85$. The SLP order parameter is also finite in this region but smaller than in the \SLPONE phase. The chirality order parameter also jumps at $h=3.8$ and decreases until $h=6.55$ at which it disappears abruptly, see Fig.~\ref{fig:hscan} inset. We interpret this as phase transitions from \SLPONE to \SLPTHREEM at $h=3.8$,
  from \SLPTHREEM into \SLPTHREECHIZERO at $h=6.55$ and from \SLPTHREECHIZERO to a disordered phase at $h=7.85$. We have repeated these mixed phase simulations for other temperatures and also for temperature scans at fixed fields to obtain the phase diagram shown in Fig.~\ref{fig:phasediagram}, where the phase boundaries are gotten from sharp jumps in the order parameters and/or energy. From the energy versus $T$ curves in the mixed phase simulations, and also Fig.~\ref{fig:pthump}, we infer that the phase transitions to the disordered state are discontinuous. No such energy discontinuities were observed between the ordered phases.}

\section{Static structure factors \label{sec:staticsf}}
The ordered SLP states cause Bragg peaks in the static structure factors, \markup{Eq.~(\ref{def_staticstructurefactor})},
%\be
%     S^{\mu \nu}(\qv) \equiv \f{1}{2} \left( S^{\mu}_{-\qv} S^{\nu}_{\qv} + S^{\nu}_{-\qv} S^{\mu}_{\qv} \right),
%\ee
which can be measured in neutron scattering experiments. \delete{$S^{\mu}_{\qv}$ is the Fourier transform of the spin $\mu$-components in the ground state.} 
\add{Using the SLP ansatz, Eq.~\eqref{eq:general ansatz}, we compute the longitudinal structure factor analytically (taking $\hv$ along the $z$-axis), giving}
\begin{align}
&S^{zz} (\qv) = \f{N^2}{32} \sum_{\Gv} \left\{ \vphantom{\f{1}{2}} \right.  \sum_{i < j} \left( u^{z2}_{ij} + v^{z2}_{ij}  \right) \Theta\! \left[\Gv \cdot (\av_i-\av_j)\right]\nonumber \\
& \times \left(  \delta_{\qv,\Gv+\Qv_{ij}} +  \delta_{\qv,-\Gv-\Qv_{ij}} \right)  \\
   &+ 2 m^{z2} \left| 1+e^{i \Gv \cdot \av_1 /2}+e^{i \Gv \cdot \av_2 /2}+e^{i \Gv \cdot \av_3 /2} \right|^2 \delta_{\qv,\Gv} \left. \vphantom{\f{1}{2}} \right\},\nonumber
\end{align}
where $ \Theta [ \Gv \cdot (\av_i-\av_j) ] \equiv   1 - \cos{(\pi (k_i-k_j))}$, $\Gv = k_1 \bv_1 + k_2 \bv_2 + k_3 \bv_3$ denotes reciprocal lattice vectors of the fcc Bravais lattice, and $k_0=0$ .
The transverse structure factor, $(S^{xx}(\qv)+S^{yy}(\qv))/2$ is similar, but with $x$- and $y$-components instead of $z$ and no $m$-term. 

\add{It follows that only the longitudinal structure factor has peaks at Brillouin zone centers $\Gamma$. In Brillouin zones where all $k_i$s are even, the $\Gamma$ peak has magnitude $\left( N m \right)^2$ reflecting the total magnetization squared, and
in Brillouin zones where $k_1+k_2+k_3$ is odd, the $\Gamma$ peak has magnitude $\left( N m/2 \right)^2$. In other Brillouin zones, there is no peak at $\Gamma$.}

\add{The $\Theta$-factor causes peaks at $\Qv_{ij}$ to occur only in Brillouin zones where $k_i - k_j$ is odd. 
We define sets of momenta
\be
\bar{Q}_{ij} \equiv  \{ \Qv_{ij} + k_1 \bv_1+ k_2 \bv_2 + k_3 \bv_3 \}, \; \; \forall  \; k_i - k_j \in \mbox{odd}.
\ee
where these peaks occur. Their peak intensities are
\begin{align}
  S^{\mu \nu}(\qv \in \pm \bar{Q}_{ij}) &=  \left( \f{N}{4} \right)^2 \left( u^{\mu}_{ij} u^{\nu}_{ij} + v^{\mu}_{ij} v^{\nu}_{ij}  \right).
\end{align}
}
 
\begin{figure}
\begin{center}
\includegraphics[width=\columnwidth]{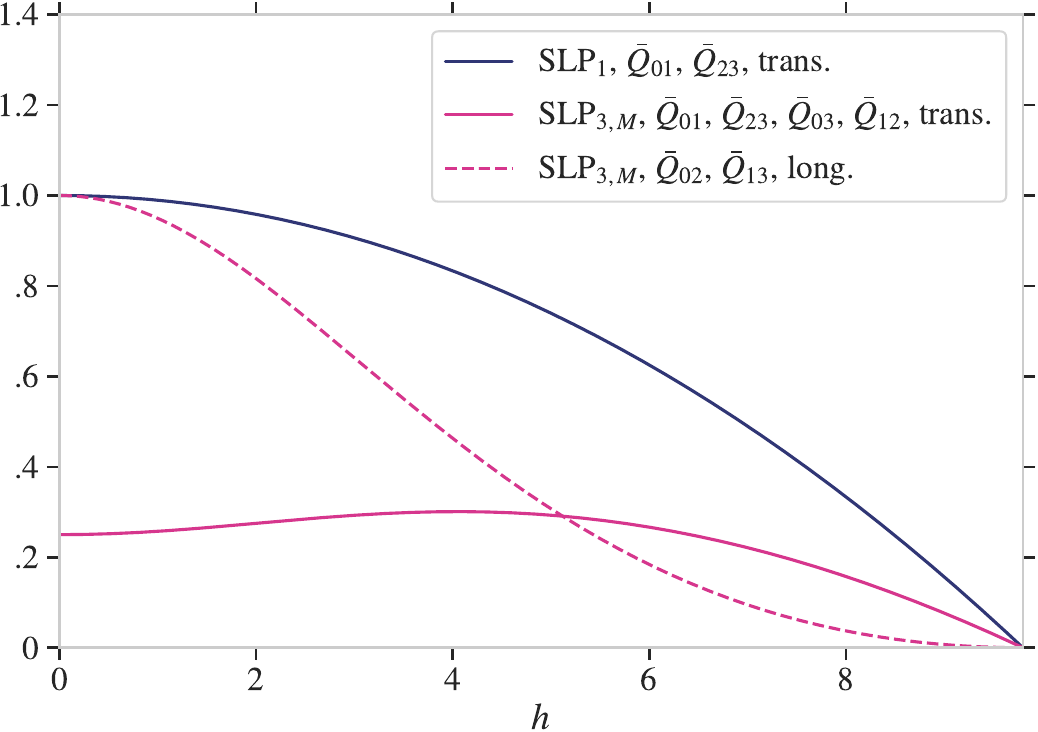}
\caption{\markup{Static structure factor peak intensities divided by $(N/4)^2$ as a function of field $h$ for the \SLPONE state and \SLPTHREEM state.}}
\label{fig:staticsf}
\end{center}
\end{figure}

\delete{
These structure factors have nontrivial peaks outside of the first Brillouin zone for $\qv = \Qv_{ij} + \Gv$ only when $k_i-k_j$ is odd.
The peak intensities are
}
%\begin{align}
%S^{\mu \nu}(\pm \Qv_{ij} + \Gv) &=  \left( \f{N}{4} \right)^2 \left( u^{\mu}_{ij} u^{\nu}_{ij} + v^{\mu}_{ij} v^{\nu}_{ij}  \right), \! \! \!  \quad k_i-k_j \in \mbox{odd}. \nonumber \\
%\end{align}
\delete{The longitudinal structure factor has an additional peak in the first Brillouin zone; at $\qv=0$ with magnitude $\left(N m \right)^2$, reflecting the total magnetization squared.
}
\add{In Fig.~\ref{fig:staticsf}, we show the static structure factors for the \SLPONE and \SLPTHREEM states as a function of field. We have not calculated it for \SLPTHREECHIZERO as we do not have an analytic expression for that state.}

\section{Dynamical structure factors \label{sec:dynamicalsf}}
From the linear spin-wave calculation outlined in Sec.~\ref{sec:spinwaveentropy} and Appendix~\ref{app:spinwaves}, we compute the zero-temperature dynamical structure factor
\be
S^{\mu \nu}(\qv,\omega) =  \int_{-\infty}^\infty \! \! \!  dt \sum_{\rv,\rvp} \langle S^\mu_{\rv}(t) S^\nu_{\rvp}(0) \rangle e^{-i \qv \cdot \left( \rv -\rvp \right)} e^{i \omega t}.
\ee
To make plots of this, we approximate the delta-functions coming from the spin-wave calculation by Gaussians centered on the excitation frequency with a standard deviation 0.01, and plot the resulting intensity \add{in the Brillouin zone centered at $\bv_1+\bv_2+\bv_3$ (111 zone) }along straight lines in momentum space going from $-{\rm K}_{ij}$ through the zone center $\Gamma$ to ${\rm K}_{ij}$ ($\Qv_{ij} = 8{\rm K}_{ij}/9$).
 
\delete{To make the discussion of locations in momentum space precise, we define sets of momenta}
%\be
%\bar{Q}_{ij} \equiv  \{ \Qv_{ij} + k_1 \bv_1+ k_2 \bv_2 + k_3 \bv_3 \}, \; \; \forall  \; k_i - k_j \in \mbox{odd}.
%\ee

We will refer to $S^{zz}(\qv,\omega)$ and $(S^{xx}(\qv,\omega)+S^{yy}(\qv,\omega))/2$ as the longitudinal and transverse dynamical structure factors, respectively.
The dynamical structure factors are in general rather complex because of the large 108-site unit cell, so we focus the discussion on the lowest energy modes.

\begin{figure*}[t]
\begin{center}
\includegraphics[width=1.6\columnwidth]{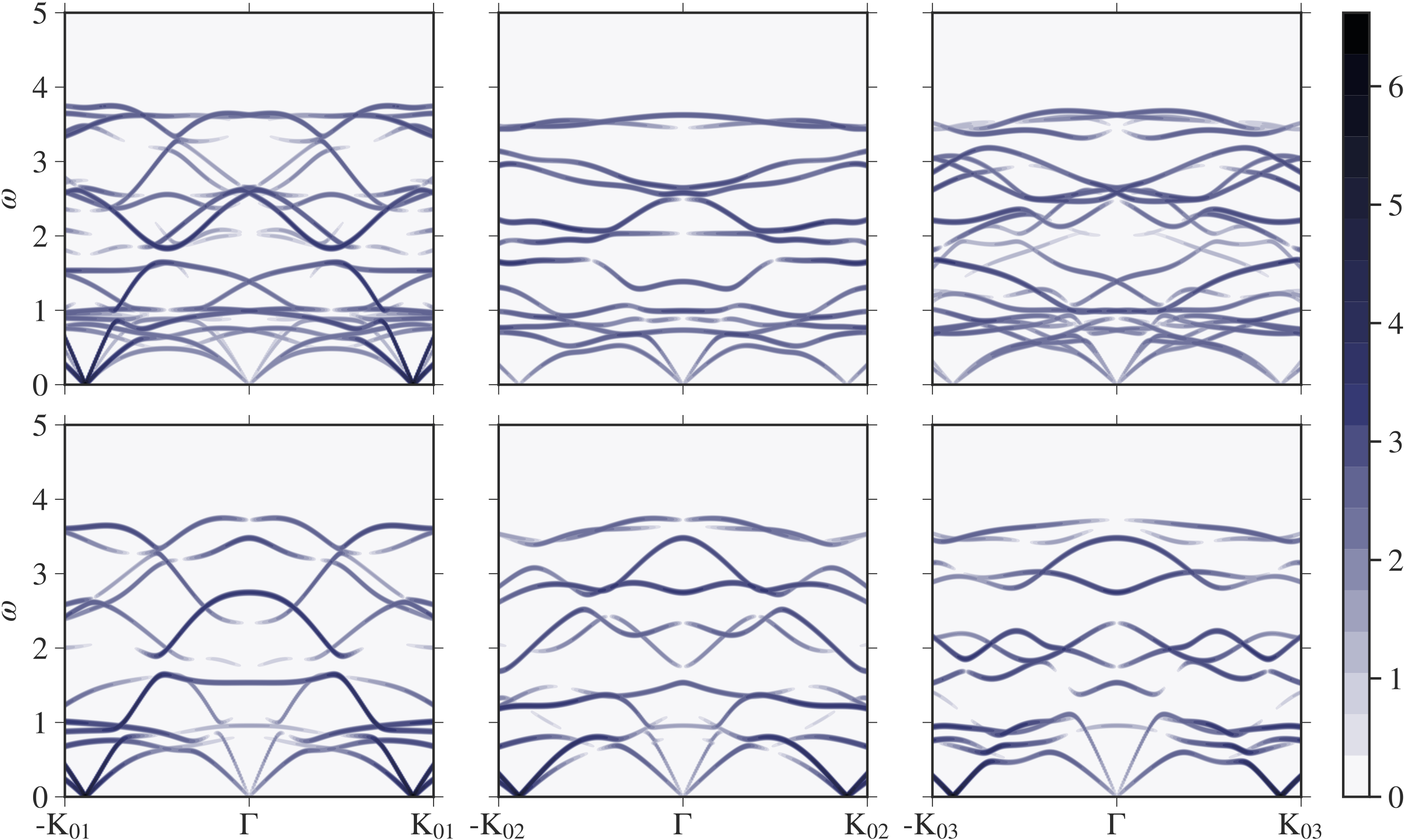}
\caption{Dynamical structure factor for the coplanar \SLPONE state in zero magnetic field in the 111 zone. Upper (lower) panels show the transverse (longitudinal) channel. The colorbar shows $\log[S(\qv,\omega)]$ with a cutoff for $S(\qv,\omega) < 1$.\label{fig:LSWalpha0h0_translong_111}}
\end{center}
\end{figure*}

In Fig.~\ref{fig:LSWalpha0h0_translong_111}, we show the transverse and longitudinal dynamical structure factors \delete{in the Brillouin zone centered at $\bv_1 + \bv_2 + \bv_3$ }for the \SLPONE state in zero magnetic field, Eq.~\eqref{eq:SLP1}, where $\uv_{01}=\uv_{23}$ and $\vv_{01}=\vv_{23}$ are chosen to lie in the $xy$-plane.
There are in total 14 different gapless modes. Two of these are associated with the transverse channel at $\bar{Q}_{01}$ and $\bar{Q}_{23}$. They connect to the Bragg peaks at the same locations caused by the explicit appearance of $\Qv_{01}$ and $\Qv_{23}$ in the ground state configuration.
The remaining 12 gapless modes are visible only in the longitudinal structure factor. They appear at all $\bar{Q}_{ij}$s, although there are no associated longitudinal Bragg peaks.
Four of these have spectral weights at $\pm \bar{Q}_{01}$ and $\pm \bar{Q}_{23}$, four at $\pm \bar{Q}_{02}$ and $\pm \bar{Q}_{13}$ and the remaining four at $\pm \bar{Q}_{03}$ and $\pm \bar{Q}_{12}$.
Slightly away from each $\bar{Q}$, however, there are at most two distinct high-intensity low-energy spin-wave branches, dependent on the directions in momentum space. We list the numbers for selected directions $\hat{D}$ in Table~\ref{tbl:numberofbranches}.

\begin{table}
\caption{
The number of longitudinal high-intensity low-energy spin-wave branches emanating out from $\bar{Q}_{ij}$ in the direction $\hat{D}$ when the ground state is \SLPONE with $\uv_{01}=\uv_{23}= \hat{x}$ and $\vv_{01}=\vv_{23}=\hat{y}$. $h=0$. $n$-fold degenerate branches are denoted: $\times n$. $^*$ refers to the directions and $\bar{Q}$s shown in Fig.~\ref{fig:LSWalpha0h0_translong_111}.
\label{tbl:numberofbranches}}
\begin{tabular}{cccc}
\toprule
$\hat{D}$                                & $(\pm \bar{Q}_{01},\pm \bar{Q}_{23})$ & $(\pm \bar{Q}_{02}, \pm \bar{Q}_{13})$  & $(\pm \bar{Q}_{03}, \pm \bar{Q}_{12})$  \\
\midrule
$\hat{K}_{01},\hat{K}_{23}$   &  $2^*$                                          &   $2 \times 2$                                         &  $1 \times 2$   \\
$\hat{K}_{02},\hat{K}_{13}$   &  $2 \times 2$                               &   $2^*$                                                           &  $1 \times 2$   \\
$\hat{K}_{03},\hat{K}_{12}$   &  2                                                 &   2                                                           &  $1^*$   \\
$\hat{X}_{1}$                          &  $2 \times 2$                                &   2                                                          &  $1 \times 2$   \\
$\hat{X}_{2}$                          &  2                                                 &   $2 \times 2$                                         &  $2 \times 2$  \\
$\hat{X}_{3}$                          &  $2 \times 2$                               &   $2 \times 2$                                         &  $1 \times 4$   \\
$\hat{L}_{0},\hat{L}_1,\hat{L}_2, \hat{L}_3 $            &  2               &   2                                                           &  1   \\
%$\hat{W}_{xpz},\hat{W}_{xmz}$ &  $2 \times 2$                           &   2                                                           &  $1 \times 2$ \\
%$\hat{W}_{xpy},\hat{W}_{xmy}$ &  2                                             &   2                                                           &  1 \\
%$\hat{W}_{xpy},\hat{W}_{xmy}$ &  2                                             &   2                                                           &  1 \\
generic                                     &  2                                             &   2                                                           &  1 \\
\bottomrule
\end{tabular}
\end{table}

The 14 gapless modes can be explained by linearizing the length constraint equations in small deviations $\delta \uv_{ij}$ and $\delta \vv_{ij}$ from the planar \SLPONE state.
There are in total 36 independent $\delta \uv_{ij}$- and $\delta \vv_{ij}$-components. In zero field, the linearized equations reduce to 22 constraints, leaving us with $36-22=14$ gapless modes.
The constraints require all $\perp$-components to be zero except for $\delta u^y_{01} = -\delta v^x_{01}$ and $\delta u^y_{23} = -\delta v^x_{23}$. The two different transverse modes are therefore caused by the relative phase between the oscillations of the unpaired sublattices (e.g. 0 and 2) being either zero or $\pi$ ("scissor" mode), where the latter has the lowest energy.
The remaining 12 modes are associated with modulations in the $z$-direction associated with each of the 12 $u^z_{ij}$'s and $v^z_{ij}$'s.

\begin{figure*}
\begin{center}
\includegraphics[width=1.6\columnwidth]{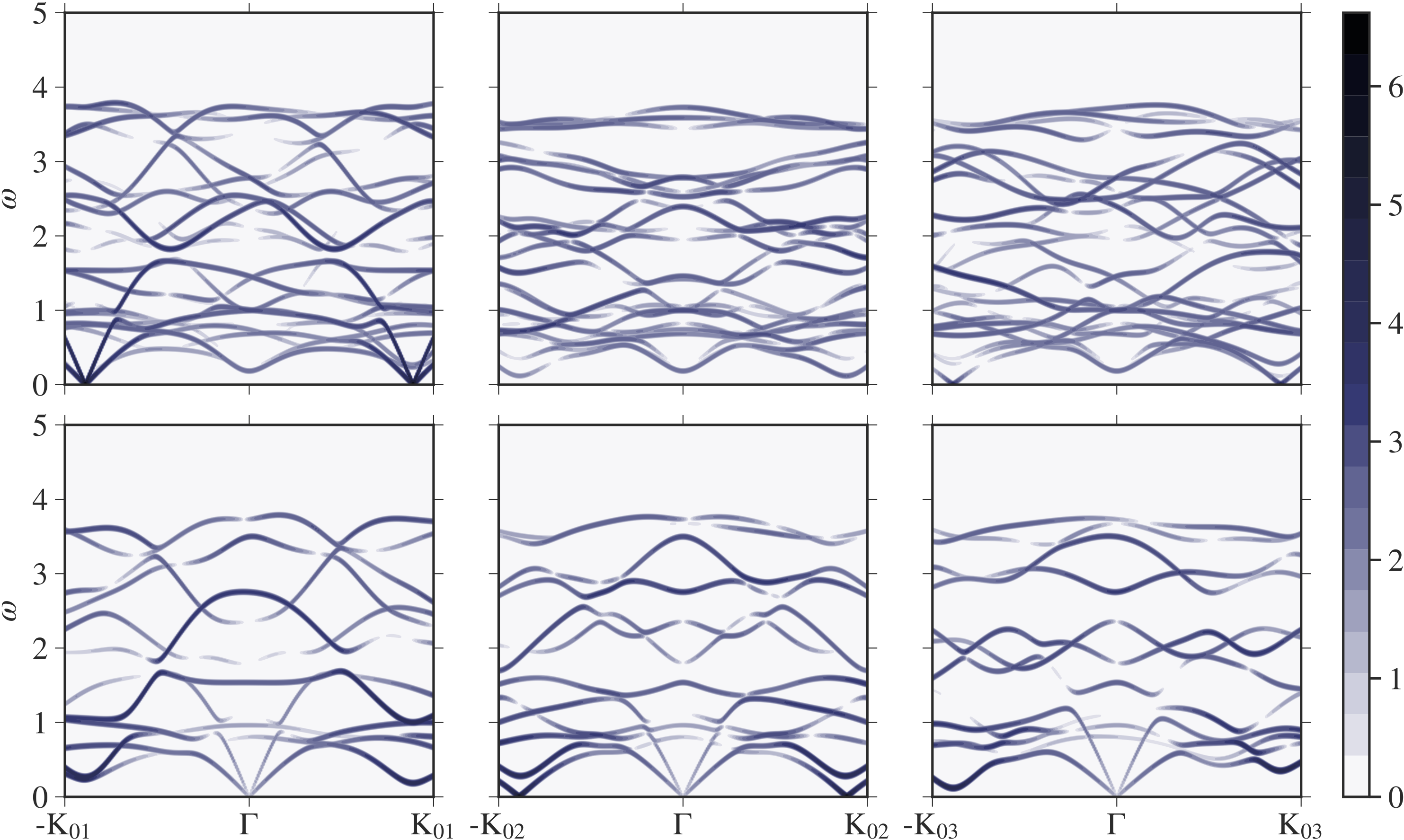}
\caption{Dynamical structure factor in the transverse (top) and longitudinal (bottom) channels for the canted \SLPONE state with $(01)\&(23)$ pairing in a magnetic field $h = 1$ in the 111 zone. The colorbar shows $\log[S(\qv,\omega)]$ with a cutoff for $S(\qv,\omega) < 1$. \label{fig:LSWalpha0h1_111}}
\end{center}
\end{figure*}

When including small magnetic fields, the \SLPONE ground state cants. In Fig.~\ref{fig:LSWalpha0h1_111}, we show the dynamical structure factor for such a canted \SLPONE state when $h = 1$.
%, corresponding to a magnetization $m \approx 0.102$.
There are four gapless modes.
Two of these are visible as two low-energy spin-wave branches near $\bar{Q}_{01}$ and $\bar{Q}_{23}$ in the transverse channel, as discussed for zero field. The remaining two modes have strong intensities both at $\bar{Q}_{02}$ and $\bar{Q}_{13}$ in the longitudinal channel, and also at $\bar{Q}_{03}$ and $\bar{Q}_{12}$ in the transverse channel. Although there are two gapless modes at these momenta, we find that only one distinct band of low-energy excitations emanate out of them; depending on the direction, either only one mode is visible, or the two are degenerate.
The four gapless modes can be explained by the four independent ways of deforming the canted \SLPONE state to linear order without energy cost, as described by the linearized length constraint equations around the canted \SLPONE solution:
\begin{align}
\delta u_{01}^y &= - \delta v_{01}^x,  \label{linearizedconstrainteq1} \\
\delta u_{23}^y &= - \delta v_{23}^x,  \label{linearizedconstrainteq2} \\
\delta u_{03}^x &= -\delta v_{03}^y = \delta u_{12}^x = \delta v_{12}^y = \tilde{m} \delta u_{13}^z =- \tilde{m} \delta u_{02}^z, \label{linearizedconstrainteq3} \\
\delta u_{03}^y &= \delta v_{03}^x = -\delta u_{12}^y = \delta v_{12}^x = \tilde{m} \delta v_{13}^z = \tilde{m} \delta v_{02}^z,  \label{linearizedconstrainteq4}
\end{align}
where $\tilde{m} \equiv m/\sqrt{1-m^2}$, and all other components $\delta \uv_{ij},\delta \vv_{ij}$ are zero.   In particular, the two last equations imply relations between the $z$-components of $\delta \uv_{02}$ and $\delta \vv_{02}$ and the $xy$-components of $\delta \uv_{03}$ and $\delta \vv_{03}$, explaining why the gapless modes at $\bar{Q}_{02}$ and $\bar{Q}_{13}$ in the longitudinal channel also show up in the transverse channel at $\bar{Q}_{03}$ and $\bar{Q}_{12}$.

\begin{figure*}[ht]
\begin{center}
  \includegraphics[width=1.6\columnwidth]{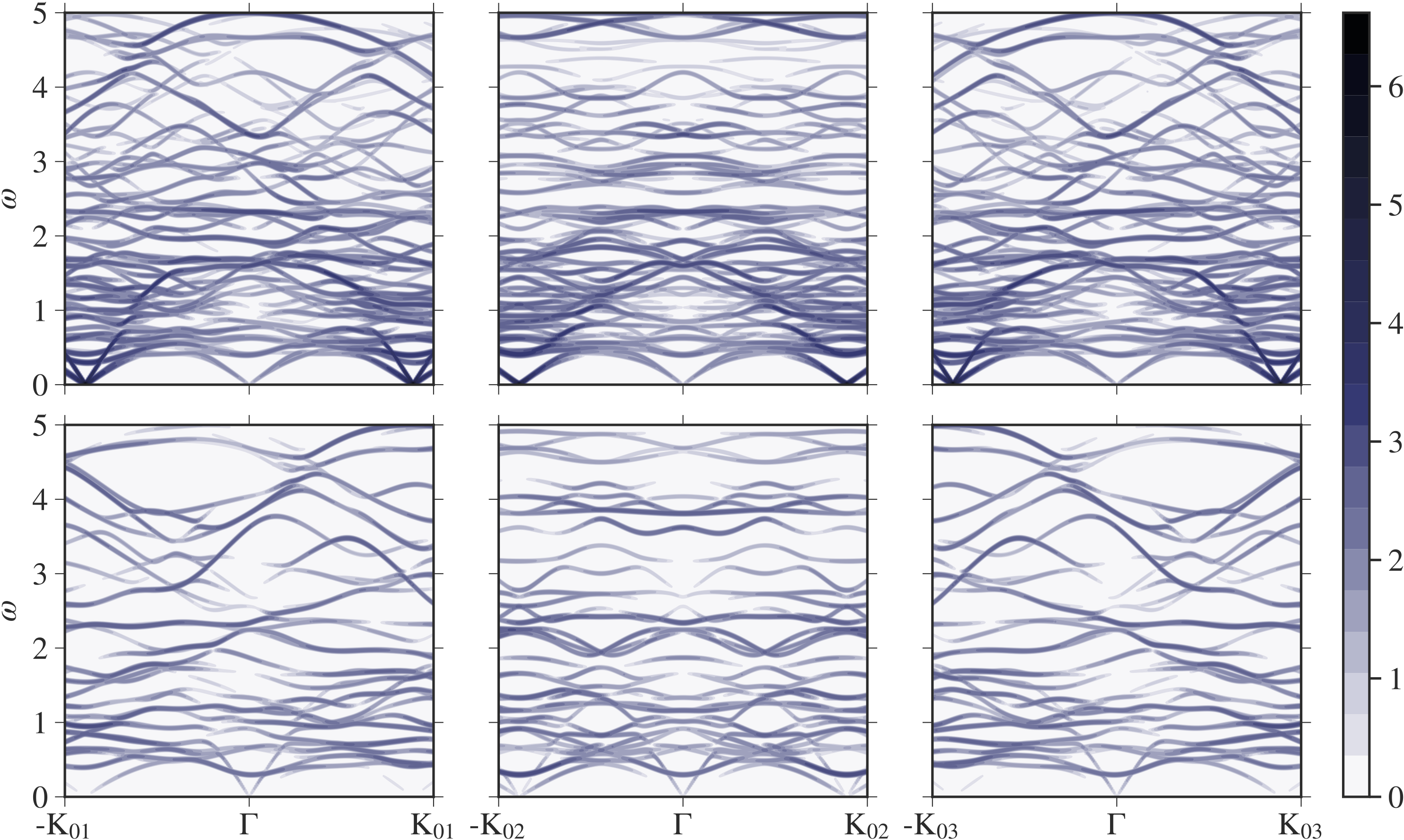}
\caption{Dynamical structure factor in the transverse (top) and longitudinal (bottom) channels for the \SLPTHREEM state in a magnetic field $h = 5$ in the 111 zone. The colorbar shows $\log[S(\qv,\omega)]$ with a cutoff for $S(\qv,\omega) < 1$.}
\label{fig:LSWalphamaxh5_111}
\end{center}
\end{figure*}

For high magnetic field ($h=5$) we calculate the dynamical structure factor for the \SLPTHREEM state; see Fig.~\ref{fig:LSWalphamaxh5_111}. It has more low-intensity spin-wave branches than \SLPONE.
We find four gapless modes that are all dominantly transverse and each one has spectral weights on eight of the 12 different $\pm \bar{Q}_{ij}$; two of the modes lack spectral weights on $\{\pm \bar{Q}_{02},\pm \bar{Q}_{13}\}$, the third on $\{ \bar{Q}_{03},-\bar{Q}_{01},-\bar{Q}_{12}, -\bar{Q}_{23} \}$, and the fourth on $\{ -\bar{Q}_{03}, \bar{Q}_{01},\bar{Q}_{12},\bar{Q}_{23} \}$. Therefore there are at most two low-energy transverse spin-wave branches emanating out of $\pm \bar{Q}_{02}$ and $\pm \bar{Q}_{13}$, and at most three for the other $\bar{Q}_{ij}$s.  Linearizing the constraint equations around the \SLPTHREEM state, we find that all $\delta u^z_{ij}=\delta v^z_{ij}=0$, which explains why the gapless modes lack spectral weight in the longitudinal channel. Furthermore the linearized equations give $\delta u_{02}^\perp=\delta u_{13}^\perp = 0$, and conditions where the in-plane components $\delta u^\perp_{ij}$ and $\delta v^\perp_{ij}$ with $ij=\{01,03,12,13,23\}$  are parametrized by the four independent components of $\delta v_{02}^{\perp}$ and $\delta v_{03}^{\perp}$.
In particular, $\delta v_{13}^{\perp}$ only depends on the two components of $\delta v_{02}^{\perp}$. From this, it follows that there are only two gapless modes at $\pm \bar{Q}_{02}$ and $\pm \bar{Q}_{13}$.

\add{The dynamical structure factors in Figs.~\ref{fig:LSWalpha0h0_translong_111}-\ref{fig:LSWalphamaxh5_111} are only shown along $-{\rm K}_{0i}\Gamma{\rm K}_{0i}$ in the 111 zone. The structure factor is identical in the 100 zone for $i = 1$, the $0\bar{1}0$ zone for $i=2$ and the 001 zone for $i=3$. Additionally, it is identical for $i=1$ along $-{\rm K}_{23}\Gamma{\rm K}_{23}$ in the 001 and 010 zones with gapless modes at $\pm\vec{Q}_{23}$ instead of $\pm\vec{Q}_{01}$, for $i=2$ along $-{\rm K}_{13}\Gamma{\rm K}_{13}$ in the $00\bar{1}$ and 100 zones with gapless modes at $\pm\vec{Q}_{13}$ and for $i=3$ along $-{\rm K}_{12}\Gamma{\rm K}_{12}$ in the $0\bar{1}0$ and $\bar{1}00$ zones with gapless modes at $\pm\vec{Q}_{12}$.  The remaining zones where $k_i-k_j$ are odd (101 and 110 for $ij = 01, 23$, 011 and 110 for $ij = 02, 13$, and 011 and 101 for $ij = 03,12$) have a generally different dynamical structure factor at high $\omega$, while the low-energy modes stay the same.}

\section{Discussion\label{sec:discussion}}
\noindent
In Ref.~\onlinecite{Glittum2023}, \SLPONE states were found to be the ground states of the classical pyrochlore Heisenberg antiferromagnet in zero magnetic field for an extended range of further-neighbor exchange interactions (with modified ordering wave vectors $\Qv_{ij}$s).  In these states, spins on pairs of sublattices form antiparallel coplanar spirals, and the ordering wave vectors for the two pairs are in general different. Here, we have found exact ground states for the pure AF $J_1$-$J_{3b}$ model in a magnetic field that generalize these \SLPONE states.

Our key finding is that the ground states can be described by the SLP ansatz in Eq.~\eqref{eq:general ansatz}. This ansatz ensures a spin sum of $4\mv$ on every tetrahedron of $J_1$ bonds, and $3\mv$ on every elementary triangle of $J_{3b}$ bonds. It therefore describes ground states for all AF couplings $J_1,J_{3b}>0$.
In the limits where either $J_{3b}=0$ or $J_{1}=0$, the SLP ansatz is however too restrictive as then {\em only} the tetrahedron condition {\em or} the triangle condition, respectively, should hold. For $J_{3b}=0$, i.e. the pure $J_1$ model on the pyrochlore lattice in a magnetic field, the ground states constitute instead an extensive manifold with one degree of freedom per tetrahedron, i.e. $N/2$ degrees of freedom in total~\cite{Penc2004}. In the opposite limit of $J_1 = 0$, the system reduces to decoupled triangular planes in a magnetic field for which the triangle condition should hold~\cite{Lee1984, Kawamura1985}. This leaves three degrees of freedom for each triangular plane. Computations of the spin-wave entropy shows that the AF $XY$ model on the triangular lattice in a magnetic field favors the coplanar Y and
$\mathbb{V}$ states~\cite{Kawamura1984}, and MC simulations have shown that the corresponding Heisenberg model orders in the same phases at finite temperature~\cite{Kawamura1985,Gvozdikova2011,Seabra2011}. The case studied here, where both $J_1$ and $J_{3b}$ are nonzero, picks out the states which the two ground state manifolds have in common.
 
Our SLP ansatz generally describes spatially periodic multi-$\Qv$ states with a large unit cell where the spins on pairs of sublattices have a tendency to be antiparallel in the plane perpendicular to the magnetic field and parallel along the field. However, the ansatz itself is incomplete as one also need\add{s} to choose its parameters such that all spins have unit length. Multi-$\Qv$ states are typically incompatible with the unit length constraint. Nevertheless, we have here presented several normalized solutions. The simplest of these are the \SLPONE states, that are identical to those found in Ref.~\onlinecite{Glittum2023} in zero magnetic field. In a finite magnetic field the coplanar \SLPONE state smoothly goes into a canted \SLPONE state that develops a uniform spin component along the magnetic field and stays almost unchanged in the plane perpendicular to the field.
This \SLPONE state is a $2\Qv$-state. The SLP ansatz harbors also $6\Qv$-states, but no normalizable $4\Qv$-states (see Appendix~\ref{app:twowavevectors}). Most interestingly, we have parametrized a family of states that interpolates continuously from the canted \SLPONE state into $6 \Qv$ \SLPTHREE states, while keeping the spins normalized.
 
Finite temperatures induce an order-by-disorder mechanism favoring the state with the largest entropy.  Using linear spin waves, we have calculated the entropy of the continuous family of SLP states. We find that for $J_1=1,J_{3b}=0.2$, the canted \SLPONE state is favored for magnetic fields $h < 4\markup{.0}$, while the 6$\Qv$-state \SLPTHREEM has the highest entropy for $h>\markup{4.0}$.
The MC simulations indicate a slightly larger value for the zero-temperature critical field of this phase transition:  $h_C \simeq 4.\markup{1}5$\add{ (see Fig.~\ref{fig:phasediagram})}.
The slight discrepancy with linear spin-wave theory is likely to be a \add{finite size effect}\delete{consequence of the relatively small systems studied}, but could also indicate a quantitative inadequacy of the linear spin-wave approximation.
\add{For the very highest magnetic fields we find that a $6\Qv$ state similar to \SLPTHREEM but with zero up tetrahedron chirality, \SLPTHREECHIZERO, is realized in the MC simulations.}
 
From the MC simulations, we conclude that the finite temperature disordering of the \SLPONE and \delete{\SLPTHREEM phases are both}\add{\SLPTHREE phases are} discontinuous. In addition, they indicate a magnetic field region (approximately \add{$h \in [2.5,9.1]$}) \delete{$h \in [2.5,4.5]$} where the system
\markup{exhibits multiple transitions
  \delete{first goes into the \SLPTHREEM phase and then into the \SLPONE phase} as the temperature is lowered in a fixed magnetic field see Fig.~\ref{fig:phasediagram}}. Lowering the field at a fixed temperature results \markup{also in multiple transitions except at the highest temperatures. 
  \delete{in two phase transitions, and even an intermediate disordered phase for the highest temperatures, see Fig.~\ref{fig:phasediagram}.}}
 
\delete{However, as evidenced by the system size dependence shown in Fig.~\ref{fig:ptdoublecv}, we expect the quantitative details of the phasediagram to change somewhat for larger system sizes.}
 
\add{We have not been able to determine the specific thermodynamic nature of the boundaries between the ordered phases in Fig.~\ref{fig:phasediagram}, as we find no thermodynamic signatures associated with them. Thus they  are likely rapid crossovers rather than true phase transitions. Nevertheless, as is evidenced by Fig.~\ref{fig:hscan}, the phases can be distinguished using a combination of order parameters and structure factors.}

The ordered SLP states should exhibit Bragg peaks in neutron scattering experiments. We have calculated the static structure factors and find Bragg peaks at momenta corresponding to the ordering wave vectors in specific Brillouin zones, see Fig.~\ref{fig:staticsf}. The discrimination between the \SLPONE and \delete{\SLPTHREEM}\add{\SLPTHREE} phases can then be made based on whether neutron scattering shows two or six inequivalent Bragg peaks at low temperatures in a single-domain single-crystalline material. Furthermore, the Bragg peaks for \SLPONE should only be visible in the transverse channel, while \SLPTHREE also shows peaks in the longitudinal channel.

We have also computed the spin-wave spectra of the SLP states and their corresponding dynamical structure factors.
As is to be expected from a spin configuration with a large unit cell, the dynamical structure factor is rather complicated, having many spin-wave branches with different intensities. Focusing on the lowest energies, the dynamical structure factors reveal gapless modes at all momenta where there is a corresponding Bragg peak in the static structure factor. This is the usual scenario where the low energy spin waves describe long wavelength deformations of a spin configuration around its ordering wave vector.
However, we also find additional high intensity gapless modes at momenta where there are no Bragg peaks.

In zero magnetic field, we find that the \SLPONE state has in total 14 different gapless modes. This is many more than guaranteed by the Goldstone theorem, where the completely broken continuous O(3) symmetry at $h=0$ leads to three gapless modes. Our results are however not in conflict with the Goldstone theorem, as it only gives a lower limit on the number of gapless modes. In fact, the number of gapless modes seen in the linear spin-wave calculation is in agreement with the number of independent solutions of the linearized spin normalization equations applied to the SLP ansatz. This indicates that the SLP ansatz is capable of describing all ground states in the vicinity of the \SLPONE state. One can speculate about the fate of the 14 gapless modes when doing a higher order spin-wave calculation. As only three are protected by symmetry, many of them will probably become gapped.

For $h \neq 0$, we find four gapless modes for both the canted \SLPONE and the \SLPTHREEM states. This is again in agreement with the number of independent solutions to the linearized normalization equations of the SLP ansatz in a field. While we expect that a higher order spin-wave calculation will gap some of the modes, it is interesting to note that a particular nonlinear deformation of the spins should be expected to change the state according to what the interpolation parameter $\alpha$ would do (see Appendix~\ref{app:explicitSLP3}), thus maintaining zero energy even for nonlinear changes.
The striking difference between the dynamical structure factors of the \SLPONE and \SLPTHREEM states is that the latter shows more intermediate intensity spin-wave branches.
Yet, for neutron scattering experiments it might be more important that \SLPONE has high-intensity low-energy {\em longitudinal} spin waves, while \SLPTHREEM only has transverse.

Higher order spin-wave calculations should be able to describe magnon decays caused by cubic bosonic terms, which generally occur for noncollinear ground states~\cite{ChernyshevZhitomirsky2006,ZhitomirskyChernyshev2013}. For a future study, it would be interesting to investigate how this affects the \SLPONE and \SLPTHREEM spin-wave spectra.

The $J_1$-$J_{3b}$ model studied here is believed to be relevant for the $\mathrm{Gd}_2B_2\mathrm{O}_7$ class of materials ($B$ is a nonmagnetic cation)~\cite{Wills2006}, however with additional dipole-dipole interactions~\cite{Raju1999,Ramirez2002,Welch2022}.
While we have not considered dipole-dipole interactions, it would be interesting to study the effects of these and other interactions on our results. We note that other Heisenberg interactions were investigated in the zero-field case in Ref.~\onlinecite{Glittum2023}. There, the \SLPONE state survives when adding $J_2$ and $J_{3a}$ as long as $J_1 > J_{3b}$ are both larger than $J_{3a}-J_2$. Particularly, it would be interesting to study the stability of the collinear SLP-X phase found close to $J_{3b} = J_{3a}-J_2$, and possible multi-$\Qv$ \SLPTHREE analogs of this phase. This would be relevant for pyrochlore spinel materials~\cite{Yaresko2008, Cheng2008} and the breathing pyrochlore materials $\mathrm{LiGaCr}_4\mathrm{O}_8$ and $\mathrm{LiInCr}_4\mathrm{O}_8$~\cite{Ghosh2019}. As most of our results rely on satisfying the tetrahedron and triangle conditions, we expect that the accurate values of $J_1$ and $J_{3b}$ are not important as long as they are both AF. Therefore, our results should apply also to breathing pyrochlores.

We hope that this work can lead to experimental searches for multi-$\Qv$ ordered phases in pyrochlore magnets, and also inspire theoretical investigations of other 3D models of frustrated magnetism.

\begin{acknowledgments}
We thank Michel Gingras for useful discussions and Claudio Castelnovo for directing us toward the $J_1$-$J_{3b}$ model.
C.G. acknowledges funding from the Aker Scholarship.
The computations were performed on resources provided by Sigma2 - the National Infrastructure for High Performance Computing and Data Storage in Norway, and on the Fox supercomputer at the University of Oslo.
\end{acknowledgments}

\section*{Data availability}

The data that support the findings of this article are openly available~\cite{data}.

\appendix

\section{Length constraints \label{app:lengthconstraints}}
In a proper classical state, each spin must have unit length.
Inserting $\Rv=n_1 \av_1 + n_2 \av_2 + n_3 \av_3$ and values for
$\Qv_{ij}$ from Table~\ref{Qijs} into the SLP ansatz Eq.~(\ref{eq:general ansatz}), we find for sublattice zero
\begin{align}
\Sv_{0,\Rv} &= \uv_{01} \cos{( \f{2\pi}{3} (n_3 - n_2))}+ \vv_{01} \sin{( \f{2\pi}{3} (n_3 - n_2))} \nonumber \\
& +\uv_{02} \cos{( \f{2\pi}{3} (n_1 - n_3))}+ \vv_{02} \sin{( \f{2\pi}{3} (n_1 - n_3))} \nonumber \\
& +\uv_{03}  \cos{( \f{2\pi}{3} (n_2 - n_1))}+ \vv_{02} \sin{( \f{2\pi}{3} (n_2 - n_1))} + \mv.\nonumber
\end{align}
Setting $k \equiv n_3-n_2$, $l \equiv n_1-n_3$ and $m \equiv n_2-n_1$, we see that $k+l+m=0$. The $2\pi/3$ factor means that the integers $k,l,m$ should be understood modulo three. Thus the SLP ansatz imply in all nine different spin possibilities on sublattice zero ($k$ and $l$ can take three independent values each, and $m$ then follows).
The nine spins in the notation $\Sv_{0,klm}$ are
\begin{align}
&\Sv_{0, 000} = \uv_{01} + \uv_{02}  + \uv_{03} + \mv, \nonumber \\
&\Sv_{0, \pm(111)} =  \f{1}{2} \left(- \uv_{01}  - \uv_{02}  - \uv_{03} +2\mv \right)   \pm \f{\sqrt{3}}{2} \left( \vv_{01} + \vv_{02} + \vv_{03} \right), \nonumber \\
&\Sv_{0, \pm(01-1)} = \f{1}{2} \left( 2\uv_{01}  -  \uv_{02}  - \uv_{03} +2\mv \right)  \pm \f{\sqrt{3}}{2} \left( \vv_{02} - \vv_{03} \right), \nonumber \\
&\Sv_{0, \pm(-101)} = \f{1}{2} \left( 2\uv_{02} - \uv_{03}  - \uv_{01} +2\mv \right)  \pm \f{\sqrt{3}}{2} \left( \vv_{03} - \vv_{01} \right), \nonumber \\
&\Sv_{0, \pm(1-10)} =  \f{1}{2} \left(2\uv_{03}  -\uv_{01} - \uv_{02}+2\mv \right) \pm \f{\sqrt{3}}{2} \left( \vv_{01} - \vv_{02} \right). \nonumber
\end{align}
Repeating this for the other sublattices, we find in total 36 length constraint equations:
\begin{align*}
 & \left( \uv_{01}+\uv_{02} +\uv_{03}  + \mv \right)^2 = 1, \\
 & \left( \uv_{01}-\uv_{12} -\uv_{13} -\mv  \right)^2 = 1,  \\
 & \left( \uv_{02} +\uv_{12} -\uv_{23}  -\mv \right)^2 = 1, \\
 & \left( \uv_{03}+\uv_{13} +\uv_{23}  - \mv \right)^2 = 1,
\end{align*}
 \begin{align*}
& \left( \uv_{01} + \uv_{02} + \uv_{03} - 2\mv \right)^2 + 3  \left( \vv_{01} + \vv_{02} +\vv_{03} \right)^2 = 4, \\
& \left( \uv_{01} - \uv_{12} - \uv_{13} + 2\mv \right)^2 + 3  \left(\vv_{01} +\vv_{12} -\vv_{13} \right)^2 = 4, \\
& \left(\uv_{02} + \uv_{12} - \uv_{23} + 2\mv \right)^2 + 3  \left( \vv_{02} + \vv_{12} +\vv_{23} \right)^2 = 4, \\
&  \left( \uv_{03} + \uv_{13} +\uv_{23} + 2\mv \right)^2 + 3  \left( \vv_{03} - \vv_{13} +\vv_{23} \right)^2 = 4,
 \end{align*}
\begin{align*}
   &  \left( 2\uv_{01} - \uv_{02} - \uv_{03} + 2\mv \right)^2 + 3 \left( \vv_{02} - \vv_{03} \right)^2 = 4, \\
&  \left(\uv_{01}-  2\uv_{02} + \uv_{03}  - 2\mv \right)^2 + 3 \left( \vv_{03} - \vv_{01} \right)^2 = 4, \\
  &  \left( \uv_{01} + \uv_{02} -2\uv_{03}  - 2\mv \right)^2 + 3\left(  \vv_{01} - \vv_{02} \right)^2 = 4,
\end{align*}
\begin{align*}
&  \left( 2\uv_{01} + \uv_{12} + \uv_{13} -2\mv \right)^2 + 3 \left( \vv_{12} + \vv_{13} \right)^2 =4,  \\
&  \left(\uv_{01}  +2\uv_{12} - \uv_{13} +2\mv \right)^2 + 3 \left( \vv_{01} + \vv_{13} \right)^2 =4,  \\
  &  \left(\uv_{01} - \uv_{12} + 2\uv_{13} +2\mv \right)^2 + 3 \left( \vv_{01} - \vv_{12} \right)^2 =4,
\end{align*}
\begin{align*}
&  \left(2\uv_{02} - \uv_{12} + \uv_{23} -2\mv \right)^2 + 3 \left( \vv_{12} - \vv_{23} \right)^2 =4,  \\
&  \left( \uv_{02} -2\uv_{12} - \uv_{23} +2\mv \right)^2 + 3 \left( \vv_{02} - \vv_{23} \right)^2 =4, \\
  &  \left( \uv_{02} + \uv_{12} +2\uv_{23} +2\mv \right)^2 + 3 \left( \vv_{02} - \vv_{12} \right)^2 =4,
\end{align*}
\begin{align*}
&  \left(2\uv_{03} - \uv_{13} - \uv_{23} -2\mv \right)^2 + 3 \left( \vv_{13} + \vv_{23} \right)^2 =4, \\
&  \left( \uv_{03} -2\uv_{13} + \uv_{23} +2\mv \right)^2 + 3 \left( \vv_{03} - \vv_{23} \right)^2 =4,  \\
  &  \left( \uv_{03}  +\uv_{13} -2\uv_{23} +2\mv \right)^2 + 3 \left( \vv_{03} + \vv_{13} \right)^2 =4,
\end{align*}
\begin{align*}
 &\left( \uv_{01}+\uv_{02}+\uv_{03} - 2\mv \right) \cdot \left( \vv_{01}+\vv_{02}+\vv_{03} \right) = 0,  \\
 & \left( \uv_{01} - \uv_{12} - \uv_{13} + 2\mv \right) \cdot \left(\vv_{01} +\vv_{12} -\vv_{13} \right) = 0, \\
& \left( \uv_{02} + \uv_{12} - \uv_{23} + 2\mv \right) \cdot \left( \vv_{02} + \vv_{12} +\vv_{23} \right) = 0, \\
 & \left( \uv_{03} + \uv_{13} +\uv_{23} + 2\mv \right) \cdot  \left( \vv_{03} - \vv_{13} +\vv_{23} \right) = 0,
\end{align*}
\begin{align*}
  &\left(2\uv_{01} - \uv_{02} - \uv_{03} +2\mv \right) \cdot \left( \vv_{02} - \vv_{03} \right) =0, \\
  &  \left( \uv_{01} -2\uv_{02} + \uv_{03} -2\mv \right) \cdot \left( \vv_{01} - \vv_{03} \right) =0, \\
  &\left( \uv_{01} +\uv_{02} -2\uv_{03} -2\mv \right) \cdot \left( \vv_{01} - \vv_{02} \right) =0,
\end{align*}
\begin{align*}
 & \left( 2\uv_{01} + \uv_{12} + \uv_{13} -2\mv \right) \cdot \left( \vv_{12} + \vv_{13} \right) =0,  \\
  &\left( \uv_{01}  +2\uv_{12} - \uv_{13} +2\mv \right) \cdot \left( \vv_{01} + \vv_{13} \right) =0, \\
  &\left( \uv_{01} - \uv_{12} + 2\uv_{13} +2\mv \right) \cdot \left( \vv_{01} - \vv_{12} \right) =0,
\end{align*}
\begin{align*}
  &\left(2\uv_{02} - \uv_{12} + \uv_{23} -2\mv \right) \cdot \left( \vv_{12} - \vv_{23} \right) =0, \\
 & \left( \uv_{02} -2\uv_{12} - \uv_{23} +2\mv \right) \cdot \left( \vv_{02} - \vv_{23} \right) =0,  \\
  &\left( \uv_{02} +\uv_{12} +2\uv_{23} +2\mv \right) \cdot \left( \vv_{02} - \vv_{12} \right) =0,
\end{align*}
\begin{align*}
  & \left(2\uv_{03} - \uv_{13} - \uv_{23} -2\mv \right) \cdot \left( \vv_{13} + \vv_{23} \right) =0,  \\
 & \left( \uv_{03} -2\uv_{13} + \uv_{23} +2\mv \right) \cdot \left( \vv_{03} - \vv_{23} \right) =0,  \\
 & \left( \uv_{03} +\uv_{13} -2\uv_{23} +2\mv \right) \cdot \left( \vv_{03} + \vv_{13} \right) =0.
\end{align*}

\section{\SLPTHREEALPHA \label{app:explicitSLP3}}
The \SLPTHREEALPHA family of ground states interpolates smoothly between \SLPONE at $\alpha=0$ and \SLPTHREEM at $\alpha = \alpha_{\rm max}$. Here, we give explicit expressions for the $\uv$'s and $\vv$'s of \SLPTHREEALPHA when the magnetic field is in the $z$-direction. We have chosen $\uv_{01}$ to be along $\hat{x}$.
With $a=(1-m^2-8m^2 \tan^2{\alpha})/2$ and $b=( (1-m^2)^2 - 16 m^2(1+3m^2) \tan^2{\alpha})^{1/2}/2$, the \SLPTHREEALPHA state is given by
\begin{align*}
  \uv_{01} &= \sqrt{a+b} \,(1,0,0), &  \vv_{01} &= |\uv_{01}| (0,-1,0),\\
  \uv_{02} &= (0,0,0), & \vv_{02} & = 4m\tan{\alpha} \, (0,0,-1), \\
  \uv_{03} &= \sqrt{a-b} \, (\sin{\alpha},\cos{\alpha},0),  & \vv_{03} &= |\uv_{03}| (-\cos{\alpha},\sin{\alpha},0),\\
  u^{xy}_{12} &= -u^{xy}_{03},  \; u^{z}_{12}= u^z_{03}, &   v^{xy}_{12} &= v^{xy}_{03},  \; v^{z}_{12}= -v^z_{03},\\
  u^{xy}_{13} &= -u^{xy}_{02},  \; u^{z}_{13}= u^{z}_{02}, &   v^{xy}_{13} &= -v^{xy}_{02},  \; v^{z}_{13}= v^z_{02},\\
  u^{xy}_{23} &=  u^{xy}_{01},  \; u^{z}_{23}= 0, &   v^{xy}_{23} &= v^{xy}_{01},  \; v^{z}_{23}= 0.
\end{align*}
Setting $\alpha=0$ one obtains the canted \SLPONE state. In the opposite limit, for $\alpha=\alpha_{\rm max}$, we get the \SLPTHREEM state:
\begin{align*}
  \uv_{01} &= \f{c_7 c_{-1}}{2c_3}(1,0,0), & \vv_{01} &= |\uv_{01}|(0,-1,0),\\
  \uv_{02} &= (0,0,0), &\vv_{02} &= \f{c_{-1}^2}{c_3}(0,0,-1),\\
  \uv_{03} &= \f{c_{-1}}{c_7} ( \f{c_{-1}^2}{2c_3}, 2m,0), &\vv_{03} &= \f{c_{-1}}{c_7} ( -2m, \f{c_{-1}^2}{2c_3},0),
\end{align*}
where we have used the abbreviations $c_{3} \equiv \sqrt{1+3m^2}$ and $c_{7} \equiv \sqrt{1+7m^2}$.

For small $\alpha$, valid to ${\cal O}(\alpha^2)$, we get
\begin{align*}
  \uv_{01} &= \left[c_{-1} - \f{4m^2 c_1^2}{c_{-1}^3} \alpha^2 \right](1,0,0), & \vv_{01} &= |\uv_{01}| (0,-1,0),\\
  \uv_{02} &= (0,0,0), & \vv_{02} &= 4m\alpha (0,0,-1), \\
  \uv_{03} &= \f{4m^2 \alpha }{c_{-1}}(\alpha,1,0), & \vv_{03} &= \f{4m^2 \alpha }{c_{-1}} (-1,\alpha,0),
\end{align*}
where we have used $c_{-1} \equiv \sqrt{1-m^2}$ and $c_1 \equiv \sqrt{1+m^2}$.

\section{Spin-wave expansion \label{app:spinwaves}}
The spin-wave expansion is carried out by first defining a rotated frame
\be
   S^{\alpha}_{\rv} = R_{\rv}^{\alpha \beta} S^{\prime \beta}_{\rv},
 \ee
such that a ferromagnetic spin configuration along the $z$-axis in the rotated frame is equal to the ground state spin configuration.
In the rotated frame, we do a Holstein-Primakoff transformation~\cite{HolsteinPrimakoff1940} to expand around the ground state
\be
\begin{pmatrix} S^{\prime x}_{\rv} \\ S^{\prime y}_{\rv} \\ S^{\prime z}_{\rv} \end{pmatrix}
=
\begin{pmatrix} \sqrt{\f{S}{2}} \left( a_{\rv}^\dagger + a_{\rv} \right) \\  i\sqrt{\f{S}{2}} \left( a_{\rv}^\dagger - a_{\rv} \right) \\  S- a^\dagger_{\rv} a_{\rv} \end{pmatrix}.
\ee
Inserting this into the Hamiltonian, the linear terms in boson operators vanish, and the quadratic part of the Hamiltonian becomes
\be
     H = \f{1}{2} \sum_{\qv} \Phi_{\qv}^\dagger H_{\qv} \Phi_{\qv},
\ee
where $\qv$ runs over the first Brillouin zone of the lattice with an unit cell of $M=4 \times 3^3=108$ spins.  $\Phi_{\qv} = [ a_{\qv,1}, \cdots , a_{\qv,M}, a^\dagger_{-\qv,1}, \cdots , a^\dagger_{-\qv,M}]^T$ is a $2M$ column vector of boson operators, and $H_{\qv}$ is a $2M \times 2M$ matrix with coefficients coming from both the rotation matrices and the parameters of the original Hamiltonian.

Being boson operators, the components of $\Phi_{\qv}$ must obey the commutation relations
\be
\left[ \Phi_{\qv,i}, \Phi^\dagger_{\qv,j} \right] = g_{ij},    \label{commrel}
\ee
where $g$ is a diagonal $2M \times 2M$ matrix with $+1$ $(-1)$ on the upper (lower) $M$ entries.
To diagonalize the Hamiltonian, we define transformed bosons $\Psi$ in the following way
\be
     \Phi_{\qv} = T_{\qv} \Psi_{\qv},
\ee
where $T_{\qv}$ is a $2M \times 2M$ matrix and $\Psi_{\qv}$ is a $2M$ column vector of transformed boson operators that also obey the commutation relations Eq.~\eqref{commrel}. This requirement leads to the following restriction on $T_{\qv}$:
\be
    T_{\qv} \, g T^\dagger_{\qv} = g.  \label{paramatrix}
\ee
 To diagonalize $H_{\qv}$ while maintaining this restriction, we follow Colpa~\cite{Colpa1978} and decompose $H_{\qv}$ using a Cholesky decomposition $H_{\qv} = K^\dagger_{\qv} K_{\qv}$. Then the Hermitian matrix $K_{\qv} g K^\dagger_{\qv}$ is diagonalized, and its eigenvalues are sorted in descending order and assigned to the diagonal matrix $D_{\qv}$. We denote by $U_{\qv}$ the corresponding unitary matrix with eigenvectors of $K_{\qv} g K^\dagger_{\qv}$ as columns.  It then follows that
 \be
      T_{\qv} = K_{\qv}^{-1} U_{\qv} \sqrt{ g D_{\qv}},
 \ee
which satisfies Eq.~\eqref{paramatrix}.
The diagonal matrix of energy eigenvalues is
\be
 E_{\qv} =  T_{\qv}^\dagger H_{\qv} T_{\qv} =   g D_{\qv}.
 \ee
 Thus, the spin-wave frequencies at $+\qv$ can be read off from the upper $M$ entries of $gD$:
 \be
 \omega_{\qv,i} = D_{\qv}^{ii}.
 \ee
This (para)diagonalization procedure fails if $H_{\qv}$ is not positive definite. Therefore, we add a small positive constant diagonal matrix to $H_{\qv}$ to keep the eigenvalues positive.

The dynamical structure factor is defined as
\be
S^{\mu \nu}(\qv,\omega) =  \int_{-\infty}^\infty \! \! \!  dt \sum_{\rv,\rvp} \langle S^\mu_{\rv}(t) S^\nu_{\rvp}(0) \rangle e^{-i \qv \cdot \left( \rv -\rvp \right)} e^{i \omega t}.
\ee
Inserting the rotation matrices and performing the Holstein-Primakoff transformation, keeping at most terms that are quadratic in the transformed boson operators, one can evaluate the boson correlators at zero temperature. Writing the transformation matrix $T_{\qv}$ on block-matrix form
\be
         T_{\qv} = \begin{pmatrix}  A_{\qv} & B_{\qv} \\ C_{\qv} & D_{\qv} \end{pmatrix}
\ee
and defining $\tilde{A}_{\qv}^{ij} \equiv e^{-i \qv \cdot \markup{\vec{\delta}_{i}}} A^{ij}$ and  $\tilde{C}_{\qv}^{ij} \equiv e^{-i \qv \cdot \markup{\vec{\delta}_{i}}} C^{ij}$, where \markup{$\vec{\delta}_{i}$} is the position of the $i$'th spin inside the unit cell, we arrive at the following expression for the dynamical structure factor
\begin{equation}
\begin{split}
& S^{\mu \nu}(\qv, \omega) = \pi NS \sum_{j=1}^{M} \delta(\omega -\omega_{\qv,j}) \\
 &  \times\sum_{i,i^\prime}^M  \left[  \tilde{A}^{\dagger j i^\prime}_{\qv}  \left( R_{i^\prime} \Lambda_1^* R^{T}_{i} \right)^{\nu \mu} \tilde{A}_{\qv}^{ij}
 +  \tilde{C}^{\dagger j i^\prime}_{\qv}  \left( R_{i^\prime} \Lambda_1 R^{T}_{i} \right)^{\nu \mu} \tilde{C}_{\qv}^{ij} \right. \\
& \left. +
\tilde{A}^{\dagger j i^\prime}_{\qv}  \left( R_{i^\prime} \Lambda_2 R^{T}_{i} \right)^{\nu \mu} \tilde{C}_{\qv}^{ij}
 +  \tilde{C}^{\dagger j i^\prime}_{\qv}  \left( R_{i^\prime} \Lambda_2^* R^{T}_{i} \right)^{\nu \mu} \tilde{A}_{\qv}^{ij}
\right], 
\end{split}
\end{equation}
where
\be
\Lambda_1 = \begin{pmatrix} 1 & i & 0 \\  -i & 1 & 0 \\ 0 & 0 &0 \end{pmatrix}, \quad
\Lambda_2 = \begin{pmatrix} 1 &  i & 0 \\  i & -1 & 0 \\ 0 & 0 &0 \end{pmatrix}. \nonumber
\ee

\section{SLP order parameter \label{app:order parameter}}
The SLP order parameter measures the staggered magnetization on chains along $\av_i-\av_j$:
\be
  \vec{C}_{\Rv_0, ij} \equiv  \f{1}{2L} \sum_{n} \left( \Sv_{\Rv_0 + n(\av_j-\av_i), i} - \Sv_{\Rv_0 + n(\av_j-\av_i), j} \right),
\ee
where $\Rv_0$ is an arbitrary up tetrahedron on the chain.
Inserting the SLP ansatz, and performing the sum over $n$, trigonometric terms with arguments $2\pi n/3$ sum to zero (for $L$ divisible by 3). The only surviving terms are those with $\Qv_{ij}$ which follows from the fact that
${\Qv_{ij} \cdot (\av_i-\av_j) = 0\mod 2\pi}$, ensuring that each term in the sum is independent of $n$. This gives
\be
  \vec{C}_{\Rv_0, ij} = \uv_{ij} \cos( \Qv_{ij} \cdot \Rv_0) + \vv_{ij} \sin( \Qv_{ij} \cdot \Rv_0).
\ee
We then square this and take the average over all parallel chains. This is achieved by summing over all points on the fcc Bravais lattice
\be
   C^2_{ij} = \f{1}{L^3} \sum_{\Rv_0}  \left( \uv_{ij} \cos( \Qv_{ij} \cdot \Rv_0) + \vv_{ij} \sin( \Qv_{ij} \cdot \Rv_0) \right)^2.
\ee
Performing this average, the trigonometric terms reduce to
\begin{align}
   &  \f{1}{L^3} \sum_{\Rv_0} \cos^2( \Qv_{ij} \cdot \Rv_0)  =  \f{1}{2}, \\
   &   \f{1}{L^3} \sum_{\Rv_0} \sin^2( \Qv_{ij} \cdot \Rv_0)  = \f{1}{2}, \\
   &    \f{1}{L^3} \sum_{\Rv_0} \cos( \Qv_{ij} \cdot \Rv_0) \sin( \Qv_{ij} \cdot \Rv_0) = 0,
\end{align}
implying that
\be
   C^2_{ij} = \f{1}{2} \left( \uv_{ij}^2  + \vv_{ij}^2 \right).
\ee
Now by considering the length constraint equations for the spins on sublattice zero, one can show that they imply
\be
\uv_{01}^2 + \vv_{01}^2 +   \uv_{02}^2 + \vv_{02}^2 +   \uv_{03}^2 + \vv_{03}^2 = 2(1-\mv^2). \label{uvs0}
\ee
Similar relations can be obtained for the other sublattices. Adding and subtracting these relations it follows that
\be
\uv_{01}^2 + \vv_{01}^2  =  \uv_{23}^2 + \vv_{23}^2,
\ee
which implies $C_{23}=C_{01}$. Similarly, $C_{13}=C_{02}$ and $C_{12}=C_{03}$. Therefore
the three component complex order parameter, which captures also other sublattice pairings
\be
O_{\rm SLP} = C_{01} C_{23} + e^{i 2\pi/3} C_{02} C_{13} +e^{i 4\pi/3} C_{03} C_{12} \nonumber
\ee
becomes
\begin{align}
 O_{\rm SLP} &= C^2_{01} + e^{i 2\pi/3} C^2_{02} +e^{i 4\pi/3} C^2_{03} \nonumber \\
              &= C^2_{01} - \f{1}{2} (C^2_{02} + C^2_{03}) + \f{\sqrt{3}}{2} i \left( C^2_{02} - C^2_{03} \right). \nonumber
 \end{align}
Squaring gives
\begin{align}
 &|O_{\rm SLP}|^2 = \left( C^2_{01} - \f{1}{2}(C^2_{02} + C^2_{03}) \right)^2 + \f{3}{4} \left( C^2_{02} - C^2_{03} \right)^2 \nonumber \\
%         &= \left( C^2_{01} \right)^2 +  \left( C^2_{02} \right)^2 +  \left( C^2_{03} \right)^2 - C^2_{01} C^2_{02} -C^2_{01} C^2_{03} -C^2_{02} C^2_{03} \nonumber \\
         &= \left( C^2_{01}  +  C^2_{02}  +  C^2_{03} \right)^2 - 3 \left( C^2_{01} C^2_{02} + C^2_{01} C^2_{03} +C^2_{02} C^2_{03} \right).\nonumber
 \end{align}
Using Eq.~\eqref{uvs0}, we get
 \begin{align}
 |O_{\rm SLP}|  &= \sqrt{ \left(1 - \mv^2 \right)^2 - 3 \left( C^2_{01} C^2_{02} + C^2_{01} C^2_{03} + C^2_{02} C^2_{03} \right)}. \nonumber
 \end{align}

\section{Two wave vectors on a sublattice \label{app:twowavevectors}}
We show in this section that the spins in an SLP state with two wave vectors on a sublattice, \SLPTWO, cannot be normalized. Setting $\uv_{03} = \vv_{03} = 0$, the nine sublattice zero equations reduce to:
\begin{align}
	\left( \uv_{01} +\uv_{02} + \mv \right)^2 &= 1, \label{2q1}\\
	\left( \uv_{01} +\uv_{02} - 2\mv \right)^2 + 3\left( \vv_{01}+\vv_{02}\right)^2 &= 4, \label{2q2}\\
     \left( - 2\uv_{01} + \uv_{02} - 2\mv \right)^2 + 3 \vv_{02} ^2 &= 4, \label{2q3}\\
     \left(      \uv_{01} - 2\uv_{02} - 2\mv \right)^2 + 3 \vv_{01}^2 &= 4, \label{2q4}\\
       \left(  \uv_{01} + \uv_{02}  - 2\mv \right)^2 + 3 \left(  \vv_{01} - \vv_{02} \right)^2 &= 4, \label{2q5}
\end{align}
\begin{align}
\left( \uv_{01} + \uv_{02} - 2\mv\right)  \cdot \vv_{01} &= 0, \label{2q6}\\
\left(\uv_{01} +\uv_{02} - 2\mv \right) \cdot \vv_{02} &= 0, \label{2q7}\\
\left( \uv_{01} -2\uv_{02} - 2\mv\right)  \cdot \vv_{01} &= 0,\label{2q8}\\
    \left( -2\uv_{01} + \uv_{02} - 2\mv\right)  \cdot\vv_{02} &= 0. \label{2q9}
\end{align}
Equations~\eqref{2q2} and~\eqref{2q5} can be combined to give
\begin{equation}
\vv_{01} \cdot \vv_{02} = 0,
\end{equation}
and the sum of Eqs.~\eqref{2q3} and~\eqref{2q4} combined with Eq.~\eqref{2q1} gives
\begin{equation}
\uv_{01} \cdot \uv_{02} = 0.
\end{equation}
If we now consider the planes spanned by these $\uv$'s and $\vv$'s, it follows from the orthogonality conditions that $\left(\uv_{01} +\uv_{02} - 2\mv \right)$ is a normal vector for the $\vv$-plane.  The remaining orthogonality conditions then give
\begin{align}
\uv_{01} \cdot \vv_{02} &= 0,\\
\uv_{02} \cdot \vv_{01} &= 0.
\end{align}
The remaining length constraints give that
\begin{align}
\uv_{01}^2 + 4\uv_{01}\cdot\mv &= \vv_{01}^2,\\
\uv_{02}^2 + 4\uv_{02}\cdot\mv &= \vv_{02}^2,
\end{align}
and
\begin{equation}
\uv_{01}^2 + \uv_{02}^2 + 2(\uv_{01} + \uv_{02})\cdot\mv = 1 - m^2.
\end{equation}
In zero magnetic field ($\mv=0$), it is consequently impossible to normalize the spins. This is because $(\uv_{01} + \uv_{02})$ is the normal vector to the $\vv$-plane (none of the vectors can have zero length, as that removes the corresponding $\Qv$ because $\uv_{01}^2 = \vv_{01}^2 $ and $\uv_{02}^2 = \vv_{02}^2$).
As $\vv_{02}$ is perpendicular to $\uv_{01}$ in addition to the $\vv$-plane normal vector $(\uv_{01} + \uv_{02})$, $\vv_{02}$ has to be normal to the $\uv$-plane. However, as $\vv_{01}$ is normal to $\vv_{02}$ it follows that $\vv_{01}$ lies in the $\uv$-plane, but that breaks the condition that $(\uv_{01} + \uv_{02})$ is normal to the $\vv$-plane.
The same conclusion holds also in the presence of a magnetic field as we have verified by direct insertion into the length constraint equations on all sublattices using Mathematica\texttrademark.

\bibliography{slp.bib}

\end{document}